\documentclass[12pt,preprint]{aastex}
\usepackage{epsf}
\usepackage{lscape}
\usepackage{graphicx}
\usepackage{natbib}
\begin{document}

\title{SHELS: A Complete Galaxy Redshift Survey with R$\leq$ 20.6}

\author {Margaret J. Geller} 
\affil{Smithsonian Astrophysical Observatory,
\\ 60 Garden St., Cambridge, MA 02138}
\email{mgeller@cfa.harvard.edu}
\author {Ho Seong Hwang} 
\affil{Smithsonian Astrophysical Observatory,
\\ 60 Garden St., Cambridge, MA 02138}
\email{hhwang@cfa.harvard.edu}
\author {Daniel G. Fabricant} 
\affil{Smithsonian Astrophysical Observatory,
\\ 60 Garden St., Cambridge, MA 02138}
\email{dfabricant@cfa.harvard.edu}
\author {Michael J. Kurtz} 
\affil{Smithsonian Astrophysical Observatory,
\\ 60 Garden St., Cambridge, MA 02138}
\email{mkurtz@cfa.harvard.edu}
\author {Ian P. Dell'Antonio}
\affil{Department of Physics, Brown University,
\\ Box 1843, Providence, RI 02912} 
\email{ian@het.brown.edu}
\author{Harus Jabran Zahid}
\affil{Institute for Astronomy, University of Hawaii at Manoa, 
\\ 2680 Woodlawn Dr., Honolulu, HI 96822}
\email{jabran@ifa.hawaii.edu}

\author{   }
\affil{   }
\email{    }

\begin{abstract}
The SHELS (Smithsonian Hectospec Lensing Survey) is a complete redshift survey covering two well-separated fields (F1 and F2) of the Deep Lens Survey to a limiting R = 20.6. Here we describe the redshift survey of the F2 field 
(R.A.$_{2000}$ = 09$^h$19$^m$32.4$^s$ and Decl.$_{2000}$ = +30$^{\circ}$00$^{\prime}$00$^{\prime\prime}$). The survey includes 16,294 new redshifts measured with the
Hectospec on the MMT.  The resulting survey of the 4 deg$^2$ F2 field is 95\% complete to R = 20.6, currently the densest survey to this magnitude limit. The median survey redshift is
$ z = 0.3$; the survey provides a view of structure in the range 0.1 $ \lesssim z \lesssim 0.6$. A movie displays the large-scale structure in the survey region. We provide a redshift, spectral index D$_n$4000, and  stellar mass for each galaxy in the survey. We also provide a metallicity for each galaxy in the range 0.2 $< z <0. 38$. To demonstrate potential applications of the survey, we examine the behavior of the index D$_n$4000 as a function of galaxy luminosity, stellar mass, and redshift. The known evolutionary and stellar mass dependent properties of the galaxy population are cleanly evident in the data. We also  show that the mass-metallicity relation previously determined from these data is robust to the analysis approach. 
\end{abstract}
\section{Introduction}
During the 1980s redshift surveys of significant portions of the nearby galaxy distribution ushered in the age of mapping the  universe (Davis et al. 1982; Geller \& Huchra 1989). Surveys with a variety of strategies rapidly explored the low redshift universe further and began to open our view of the intermediate redshift universe (e.g.  Rowan Robinson et al. 1990; Lilly et al. 1995; Shectman et al. 1996; Yee et al. 1996; da Costa et al. 1998). 
Le F\'evre et al. (2013) summarize  (their Figures 24 through 26 and associated references) the characteristics of the remarkable array of redshift surveys that now cover enormous volumes nearby and  probe the universe to large redshift. 
 
The Sloan Digital Sky Survey (SDSS hereafter; Ahn et al. 2014), large-area southern surveys to comparable or somewhat greater depth (e.g. Shectman et al. 1996; Colless et al. 2001; Jones et al. 2009: Baldry et al. 2010), and probes of the distant universe have been the basis for revisions and new tests of our understanding of the growth of structure in the universe. However, largely as a result of the available instrument/telescope combinations, there are few dense surveys covering the redshift range $ 0.2 < z < 0.6$ over significant areas.

Here we describe SHELS (Smithsonian Hectospec Redshift Survey), a dense redshift survey covering 8 deg$^2$ in two well-separated 4 deg$^2$ fields; the survey is essentially complete to a limiting R = 20.6. The survey takes its name from the original goal of comparing the matter distribution revealed by a foreground redshift survey with the weak lensing maps of the Deep Lensing Survey (DLS hereafter; Wittman et al. 2006). SHELS covers two of the five 4 deg$^2$ fields of the DLS, F1 and F2. We discuss the survey of F2 here and plan to report later on the intrinsically less dense survey field, F1 (the redshift survey is not yet complete). When complete the two well-separated regions should provide an improved measure of the impact of cosmic variance
relative to contiguous field covering the same area on the sky. Taken together the two SHELS fields will include $\sim 21,500$ redshifts for galaxies with R $\leq$ 20.6; F2 alone includes 12,705 redshifts (reported here) to this limit. The median redshift $z = 0.3$. The SHELS F2 field is 95\% complete to R=20.6 and is currently the most densely sampled region to this limit.

The AGES survey (Kochanek et al. 2012) focuses on AGN evolution and the galaxy redshift survey covers an area and redshift range most comparable with SHELS. Kochanek et al. (2012) also used the
300-fiber Hectospec (Fabricant et al. 2005) on the MMT to measure redshifts. Their galaxy redshift survey covers 7.7 deg$^2$ and includes galaxies with I$< 20$; the survey includes AGN to fainter
limits. The median redshift is $z = 0.31$ nearly identical to SHELS.
AGES uses a complex sparse sampling strategy in several photometric bands. To the SHELS limit, R = 20.6, the number density of AGES redshifts is $\sim$1600 deg$^{-2}$
whereas the SHELS density averaged over the two fields is $\sim 2700$ redshifts deg$^{-2}$.
Thus SHELS, which is essentially complete to its magnitude limit, complements AGES by providing a sample with a simple selection in apparent magnitude.  

We have used subsets of the SHELS data for a variety of projects that take advantage of
its straightforward selection. As initially intended, we 
have compared the matter distribution in the universe marked by galaxies in the redshift survey with the projected distribution measure with a weak lensing map (Geller et al. 2005; Geller et al. 2010). Cross-correlation of the two maps shows that the lensing map images the matter distribution traced by galaxies in the redshift survey (Geller et al. 2005). Comparison of clusters of galaxies identified in the redshift survey and from the weak lensing map suggests that the threshold for cluster detection in weak lensing maps should be more conservative to yield reliable samples of real systems (Geller et al. 2010; Utsumi et al. 2014).

The uniform spectroscopy for the SHELS survey provided a basis for computation of the H$\alpha$ luminosity function (Westra et al. 2010) and for a determination of the faint end of the galaxy luminosity function at $ z \leq 0.1$ (Geller et al. 2012). We have also used SHELS to explore the nature of star-forming galaxies detected with WISE (Wide-Field Infra-Red Explorer; Wright et al. 2010) at 22$\mu$m (Hwang et al. 2012b) and to measure the mass-metallicity relation for galaxies in the redshift range $0.2 < z < 0.38$ (Zahid et al. 2013). Here we provide the metallicities and stellar masses for galaxies in this redshift range.

The SHELS data have also provided a partial basis for calibration of spectral indicators determined from Hectospec spectra (Fabricant et al. 2008). The data have also been part of the foundation for a demonstration that central velocity dispersions can be measured reliably with the Hectospec (Fabricant et al. 2013). We plan to report central velocity dispersions for galaxies in both the F1 and F2 DLS fields in a separate paper.

We describe the data in Section \ref{Sdata}. We include the definition we use for
the spectral index D$_n$4000, the way we determine the stellar mass for each galaxy, and the approach to measuring the metallicity for star-forming galaxies in the redshift range $0.2 < z < 0.38$. We describe the survey completeness in Section \ref{Scomplete}. We discuss both the observed and rest-frame color distribution for the surey galaxies in Section \ref{Scolor}. To provide an astrophysical context and a partial guide to potential uses of the data we provide here, we briefly review the way previous workers have used the indicator D$_n$4000 as
an evolutionary marker in Section \ref{Smarkers}. Then in Sections \ref{Sbreak} and \ref{Smass}, we show our measured distributions of D$_n$4000 as functions of redshift, intrinsic galaxy luminosity, and stellar mass. We demonstrate  that the salient evolutionary effects are qualitatively evident in the SHELS F2 data. We also provide a set of spectra in bins of stellar mass and redshift. Finally in Section \ref{SMZrel} we reprise the study of Zahid et al. (2013) and show that the mass-metallicity relation can be recovered from summed spectra. We conclude in Section \ref{Sconc}. 

We adopt H$_0$ = 70 km s$^{-1}$ Mpc$^{-1}$, $\Omega_\Lambda$ = 0.7 and $\Omega_m$ = 0.3 throughout.
All quoted errors in measured quantities are 1$\sigma$.

\section {The Data}
\label{Sdata}
The SHELS redshift survey covers two of the five fields of the ambitious
Deep Lens Survey (DLS; Wittman et al. 2006); these fields are denoted F1 and F2 and each covers $\sim$4 square degrees. Our goal is to achieve a redshift survey of each field with a completeness of $\gtrsim$90\% to the magnitude limit, R$\sim 20.6$. This level of completeness is comparable with the
SDSS Main Galaxy Sample with limiting r = 17.77 (Strauss et al. 2002; Park \& Hwang 2009).

The F1 field is centered at R.A.$_{2000}$ = 00$^h$53$^m$25.3$^s$ and
Decl.$_{2000}$ = 12$^\circ$33$^{\prime}$55$^{\prime\prime}$. The redshift survey of this field is not yet complete, but there are $\sim$ 9800 extended sources with R$\leq 20.6$ in this field. The redshift survey of this field will include $\gtrsim 8800$ galaxies.

Here we describe the redshift survey of the F2 field centered at 
R.A.$_{2000}$ = 09$^h$19$^m$32.4$^s$ and 
Decl.$_{2000}$ = +30$^{\circ}$00$^{\prime}$00$^{\prime\prime}$. For this field we base our redshift survey on the DLS photometry. The DLS photometric data, 
with an effective exposure time of 14,500 seconds on the Mayall 4-meter in $<0.9^{\prime\prime}$ seeing, reach a 1 $\sigma$  limit for source detection in R of 28.7 magnitudes arcsec$^{-2}$ (Muller et al 1998; Wittman et al 2006).

The redshift survey (Geller et al. 2005; Geller et al. 2010; Westra et al. 2010; Geller et al. 2012; Hwang et al. 2012b)
covers the F2 field to a limiting apparent magnitude 
R = 20.6. In the complete survey region of F2 with R$\leq$ 20.6, 12,705 galaxies have a spectroscopically determined redshift and the survey is 95\% complete to the magnitude limit. Ultimately the SHELS
survey of the F1 and F2 fields will include $\gtrsim$ 21,500 galaxies with R $\leq$ 20.6; SHELS is currently the densest
survey covering the redshift range $0.2 \lesssim z \lesssim 0.6$.

\subsection {Photometry}
\label{photometry}

The DLS selected all five of their fields, including F2, to 
exclude apparently bright nearby galaxies and to avoid known rich clusters with
redshift $z \lesssim 0.1$. The F2 field is distinctive because it contains a complex of rich clusters at $z \simeq 0.3$; these systems include Abell 781 (Abell 1958).   

Wittman et al. (2006) describe the photometric data reduction pipeline for the DLS.
The R band source list for the F2 field is the basis for our galaxy catalog. 
The automatic object identification algorithm
produces a complete F2 catalog of objects with surface brightness $\mu_{50,R} \leq 27.0$ within the half-light radius. The DLS galaxy magnitudes we list are extrapolated total magnitudes; they are extrapolated from isophotal
magnitudes within the limiting 28.7 mag arcsec$^{-2}$ isophote.

To construct the catalog for SHELS spectroscopy, we examined all of the objects with 
R$  \leq 20.6$ visually to remove the obvious artifacts. We conservatively included some apparently stellar objects in the observing list. In fact, among the objects  we observed spectroscopically, 
$\sim 3$\%  are stars with a median apparent magnitude R$\sim$19.3. At the low surface brightness extreme of the data, most objects we inspected were obvious artifacts; we included all
the low surface brightness galaxy candidates as spectroscopic targets. Geller et al.
(2012) describe the completeness of the galaxy catalog as a function of surface brightness. 

Because of the extended (power-law) wings of the PSF, regions of the sky near bright ($R<15$) stars produce contaminated catalogs where detected objects have incorrect photometry and size estimates. Brighter stars affect a larger area, but effectively measuring photometry for these bright stars in the DLS images is impossible. 
We thus develop an empirically calibrated measurement of the size of the affected area based on the USNO-A2 (Monet et al. 2003) m$_V$ magnitudes for stars.
We measured the regions with significant ($>1\sigma$) background enhancement around bright stars in one subfield of F2 (the p11 subfield), and compared the radius of the region with the USNO-A2 V magnitude to develop an empirical relation for the exclusion regions.  The best-fitting relation is:
\begin{equation}
 r_{exc} = 1.2\left(m_V-15.5\right)^2 {\rm arcsec}
\end{equation}

\noindent where $r_{exc}$ is the radius of the excluded region.

For the few brightest stars, we had to adjust the size of the region based on the color of the star (not surprising given the central wavelength difference between the DLS R filter and the photographic V filter used in USNO).  For most stars, however, we use the exclusion region specified by the USNO magnitude.

In addition to the power-law PSF ``halos'', very bright stars have significant bleed trails in the DLS images.  Although these trails constitute a smaller portion of the image, they also contaminate the photometry of galaxies near them.  Using the SDSS, we recovered photometry for a few  very bright galaxies with saturated cores and for 252 objects with isophotes overlapping a bleed trail but within the complete survey region.

\begin{figure}
\centerline{\includegraphics[width=7.0in]{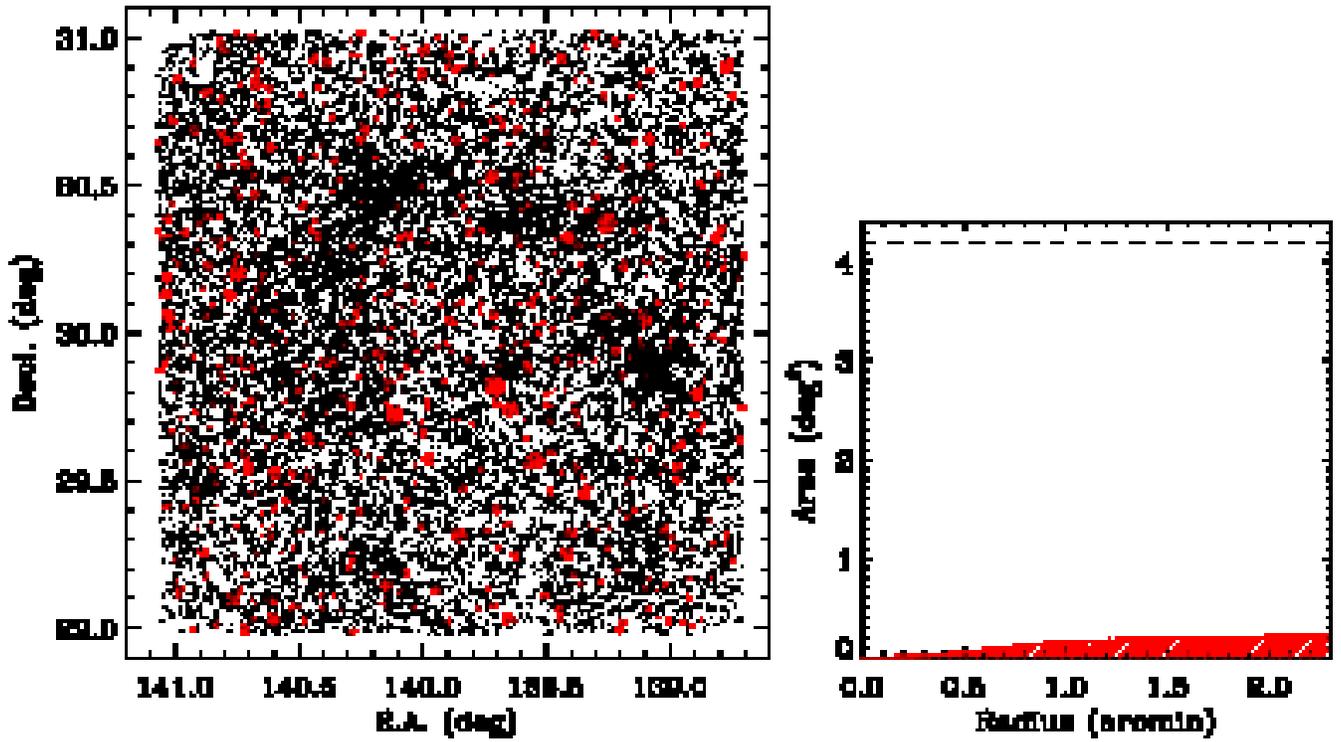}}
\vskip 5ex
\caption{Survey mask (left) along with the cumulative area covered by the masked regions as a function of angular radius of the region (right). The red dots in the left-hand panel show the size of the masked regions; the dots indicate galaxies in the survey with measured redshifts.
\label{Fmask}}
\end{figure}

Table \ref{tab-mask} lists the central coordinate and the radius for each of the masked regions.  Figure \ref{Fmask} shows the distribution of the masked regions superimposed
on the survey region (left panel) and the cumulative area they cover as a function of their radii (right panel).  The total area of the DLS F2 field is 4.19 deg$^2$. The masked regions cover 0.21 deg$^2$, or about 5\% of the area. The complete redshift survey covers the unmasked region of 3.98 deg$^2$.

We included all 13,249 galaxy candidates with Kron-Cousins R $\leq $ 20.6 and within the unmasked region as targets for spectroscopic observation. 
All but 70 of the galaxy candidates with R$\leq 20.6$ in DLS F2 catalog also have SDSS photometry. These 70 objects are generally unresolved in the SDSS.

\subsection {Spectroscopy}
\label{spectroscopy}

We acquired spectra for the objects with the Hectospec 
(Fabricant et al. 1998, 2005) on the MMT from
April 13, 2004 to December 21, 2009. The instrument deploys 300 fibers over a 1$^\circ$
field. The Hectospec observation
planning software (Roll et al. 1998) 
enables  efficient sample acquisition. We filled unused survey object fibers with  targets
fainter than the survey limit.

The SHELS spectra cover the wavelength range 3,700 --- 9,100 \AA\ with a resolution of $\sim$5 \AA.\ Exposure times ranged from 0.75 to 2 hours. We reduced the data with the standard Hectospec pipeline
(Mink et al. 2007) and derived redshifts with RVSAO (Kurtz \& Mink 1998) with
templates constructed for this purpose (Fabricant et al. 2005). Our 1651 pairs of observations for absorption-line objects imply a mean internal error (normalized by $(1+z)$) of 48 km s$^{-1}$; for the 238 pairs of emission-line objects, the mean internal error is 24 km$^{-1}$. The comparison pairs have redshifts $\leq$ 0.7. For 427 overlaps between our observations and the SDSS, the rms difference between the Hectospec and SDSS redshift measurements is 37 km s$^{-1}$ (also normalized by $(1+z)$), but these objects are much brighter than the sample limit. 

During the pipeline processing, we visually inspect all of the spectra and
assign a quality flag of ``Q'' for high-quality redshifts, ``?'' for
marginal cases, or ``X'' for poor quality to the redshift determination. We report only high quality ``Q'' redshifts.

\begin{figure}
\centerline{\includegraphics[width=7.0in]{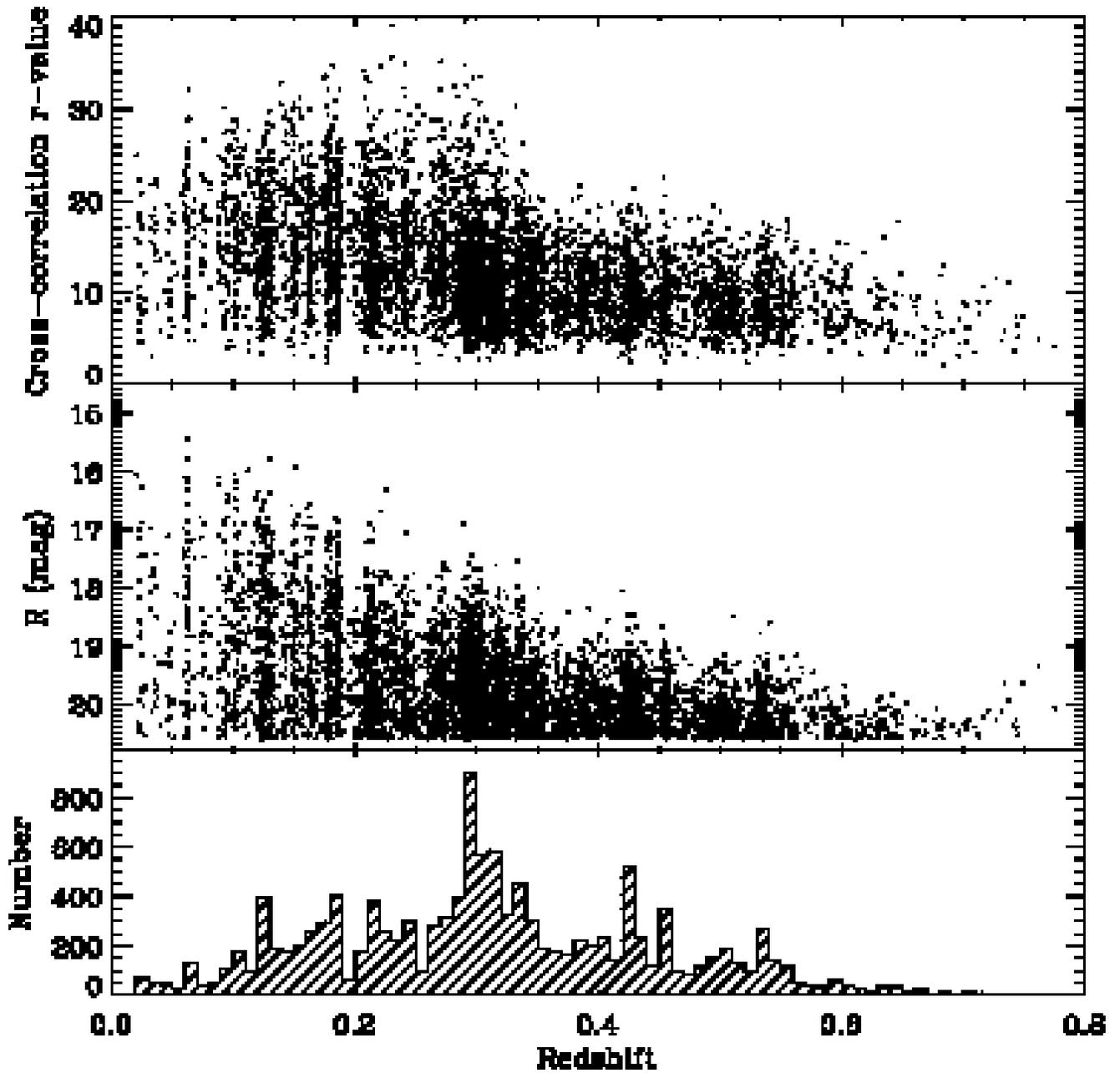}}
\vskip 5ex
\caption{Cross-correlation $r$-value (Tonry \& Davis 1979), a redshift  quality indicator, as a function of redshift (upper panel). The center panel shows apparent R-band magnitude as a function of redshift. Points scattered above the faint limit result from
poor photometry. The lower panel shows a redshift histogram in bins of 
$\Delta{z} = 0.01$. 
\label{Fzxcr}}
\end{figure}

To give a view of the quality of the spectroscopy, Figure \ref{Fzxcr} shows the distribution of the Tonry \& Davis (1979) $r_{TD}$ statistic as a function of redshift. The statistic $r_{TD}$ measures the relative amplitude of the cross-correlation peak and thus 
provides an estimate of the error in the redshift. We compute the error in the redshift according to the prescription in Kurtz \& Mink (1998). The maximum $r_{TD}$ value declines for apparently fainter galaxies at large redshift. The scatter of values with $r_{TD} \lesssim 5$ results from the
visual identification of reliable features in otherwise noisy spectra. Fabricant et al. (2005) discuss the Hectospec data reduction in more detail.

\begin{figure}
\centerline{\includegraphics[width=7.0in]{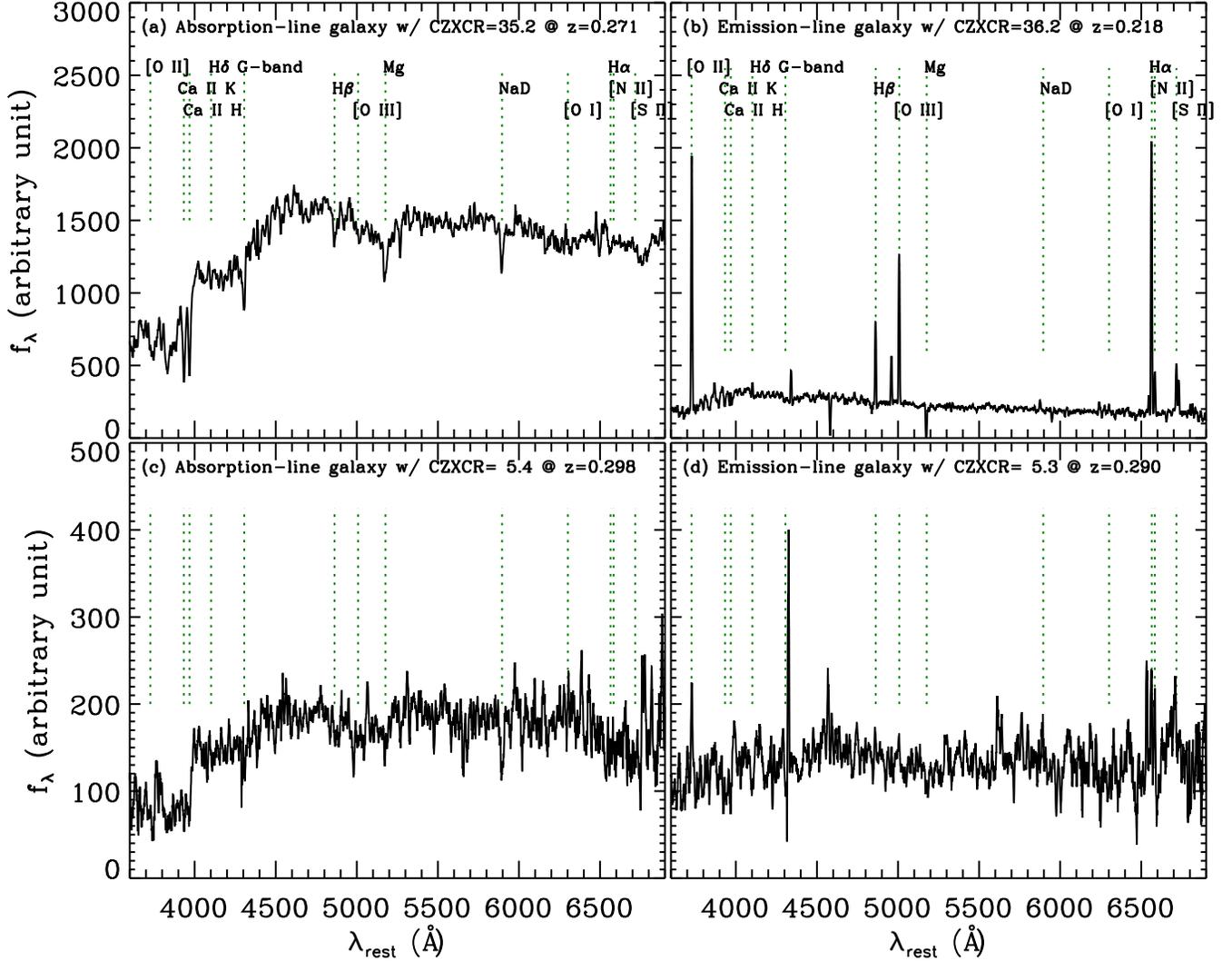}}
\vskip 5ex
\caption{Sample absorption-line (left) and emission-line spectra (right) demonstrating the 
range of quality (cross-correlation coefficient) at $z \sim 0.29$. Labels indicate 
major spectral features; unlabeled spikes are
badly subtracted night sky lines.  
\label{Fegsp}}
\end{figure}

Figure \ref{Fegsp} shows four sample spectra: there are two absorption-line galaxies and two emission-line objects. We show examples for high and low values of the cross-correlation measure $r_{TD}$. 

The F2 redshift survey in the complete region (unmasked)  includes 12,705 galaxies with a measured redshift and with R $\leq$ 20.6. Among these galaxies, 25 have a redshift from the SDSS; all of the rest are from our MMT Hectospec observations.
The integral completeness of the redshift survey to this limit is 95\%.

Geometric constraints are responsible for many of the  708 objects 
without a redshift; a majority are near the survey 
corners and edges. On average, Hectospec positionings revisit every region within the DLS field more than a dozen times. We are thus minimally biased against galaxies with nearby neighbors.

Table \ref{tbl:props} lists the number of galaxies in the complete photometric samples and the number of objects with a redshift. 
Table \ref{tab-samp} lists the Hectospec (and SDSS) redshifts for 13,325 galaxies with R$\leq$ 20.6. The table includes the SHELS ID, the SDSS ObjID, the total R magnitude from the DLS photometry
along with its error, the redshift and its formal error 
derived from the $r_{TD}$ value, the redshift source, and a flag indicating whether the object is in the masked region.   Objects designated 1 are within the masked region and we do not include these objects when we analyze the data. Their redshifts may obviously be useful for some other purposes, but their DLS photometry is unreliable and completeness is difficult to evaluate in these regions. Table \ref{tab-samp} also includes derived quantities that we have used for analysis here or in previous papers. These quantities are D$_n$4000, the stellar mass and its error, and the metallicity (for emission-line galaxies with 0.2$ < z < 0.38)$.  

We provide a value of D$_n$4000 for each galaxy in the completely surveyed region (unmasked). We do not report D$_n$4000 $<$ 0 or D$_n$ 4000$>$ 3; these values result from poor quality spectra that are just adequate to yield a redshift. For each galaxy we also provide a stellar mass provided that the procedure outlined in Section \ref {Sspecprops} converges.

Table \ref{tab-sampfa} lists 2994 redshifts for galaxies with $R > 20.6$ in the F2 DLS field. This set of objects is not complete in any way, but these redshifts may be a useful complement to other datasets. All of these objects have SDSS counterparts. 
Finally Table \ref{tab-samp_noz} lists the 703 galaxy candidates that complete the photometric survey in the unmasked reegion of F2: they have R$\leq$20.6 but no redshift.

\subsection {The D$_n$4000, Stellar Mass, and Metallicity for the F2 Field} 
\label{Sspecprops}

As in Fabricant et al. (2008), we define D$_n$4000  as the ratio of flux (in $f_\nu$ units)
in the 4000-4100 \AA\ and 3850-3950 \AA\ bands (Balogh et al. 1999). D$_n$4000 is a measure of the amplitude of the 4000 \AA\ break in the spectrum and provides a coarse estimate of the aggregate population age (although entwined with the metallicity). The error in D$_n$4000, based on 1468 repeat measurements,  is 0.045 times the value of the index (Fabricant et al. 2008). Fabricant et al. (2008) also use 358 galaxies in common with the SDSS to demonstrate that the D$_n$4000 we measure from a Hectospec spectrum agrees with the SDSS
determination.

As in Zahid et al. (2013), we derive stellar masses by applying the 
Le Phare\footnote{http://www.cfht.hawaii.edu/$\sim$arnouts/LEPHARE/lephare.html} code written by Arnouts \& Ilbert to SDSS five-band photometry. Fits of stellar population models to the spectral energy distribution provide a mass-to-light ratio. We scale the luminosity by the mass-to-light ratio to obtain the stellar mass (Bell et al. 2003). The set of models (based on the stellar templates of Bruzual \& Charlot (2003) and the Chabrier (2003) IMF) includes two metallicities 
(0.4 and 1 solar) and seven exponentially decreasing star formation rates. We compute models with a star formation rate  SFR$\propto$e$^{-t/\tau}$ for $\tau$ = 0.1, 0.3, 1.2, 3, 5, 10, 15, and 30 Gyr. We allow $0 <$ E(B-V)$ < 0.6$ following the Calzetti et al. (2000) extinction law. Stellar population ages range from 0 to 13 Gyr, but the age never exceeds the age of the universe at the galaxy redshift. For each object, the stellar mass is the median of the model values. 

We compare our stellar mass estimates to the SDSS as described by Zahid et al. (2013). There is a systematic offset of 0.15 dex between our values and those derived by the MPA/JHU group. After correction for the offset there is a 0.17 dex dispersion between our estimates and the MPA/JHU values, consistent with the expected errors in the two techniques. In our tables we list stellar masses directly from the Le Phare routines.

Again as in Zahid et al. (2013), we use the R23 line ratio calibrated by Kobulnicky \& Kewley (2004; KK04) to derive metallicities for galaxies in the redshift
range 0.2$ < z < 0.38$. As in Zahid et al. (2013) we restrict the metallicity measurement to this range because H$\alpha$ is outside the Hectospec bandpass for $z \geq 0.38$ and because aperture effects may be large for $ z \lesssim$ 0.2 (Kewley et al. 2005). 

The indicator R23 we use is defined by ratios of line intensities:
\begin{equation}
R23 = {{[OII]\lambda3727 +[OIII]\lambda4959,5007}\over{H\beta}}
\end {equation}
where we assume the recombination value, 3, for the ratio of the [OIII]$\lambda$5007 to 
[OIII]$\lambda$4959 line strengths. We use the ratio
\begin{equation}
O23 = {{[OIII]\lambda4959,5007}\over{[OII]\lambda3727}}
\end{equation}
to correct for variation in the ionization state of the gas. 

We  use the H$\alpha$ and [NII]$\lambda$6584 lines to select star-forming galaxies. We also require a S/N$ > 3$ for these lines as well as for H$\beta$ and [OII]$\lambda$3727. We did not apply a S/N cut on [OIII] because Foster et al. (2012) show that such a cut can lead to a significant bias in the determination of the mass-metallicity relation. Zahid et al. (2013) discuss the mass-metallicity relation based in part on these data. In Section \ref{SMZrel} we display summed spectra that demonstrate the salient features of the relation.

\subsection {Redshift Survey Completeness}
\label{Scomplete}
 
\begin{figure}
\centerline{\includegraphics[width=7.0in]{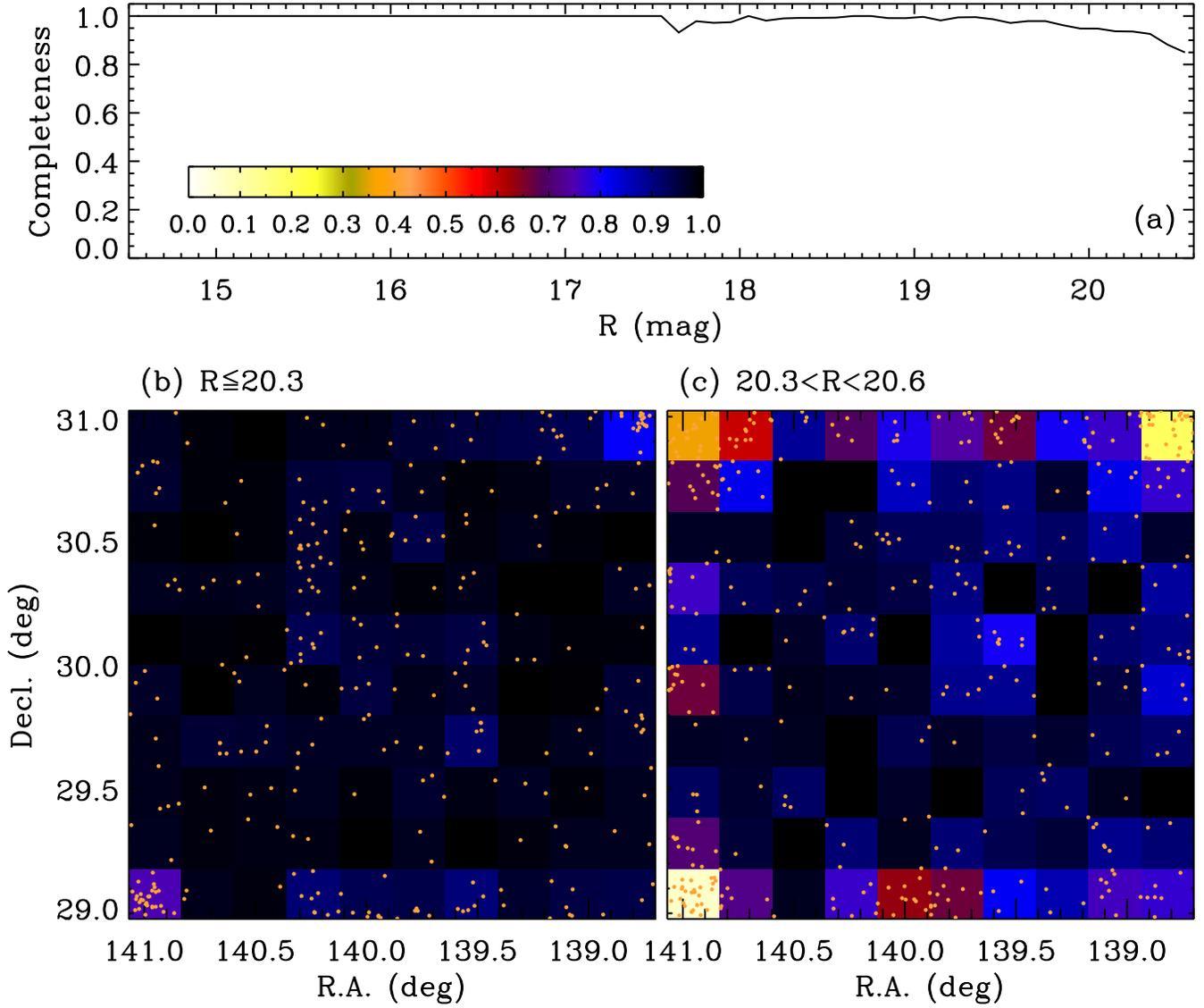}}
\vskip 5ex
\caption{Completeness of the SHELS redshift survey of the DLS F2 field. The upper panel shows the completeness as a function of DLS R-band magnitude. The color bar insert shows the completeness fractions for the spatial completeness displays in the lower two panels.
The lower left panel shows the completeness in 12$\times$12 arcminute bins for 
galaxies with R $ < 20.3$. The yellow points indicate galaxies in the photometric sample without a measured redshift. The right hand plot shows the completeness in the interval 20.3$ <$R$<20.6$ in the same format. Note that the only significant incompleteness occurs at the
corners and edges of the field.
 \label{Fcomplete}}
\end{figure}

Figure \ref{Fcomplete} shows the differential completeness of the survey as a function of  the limiting R-band magnitude (upper panel). The completeness declines  very slowly to R$\simeq$ 20.3 and then declines a bit more steeply to the limiting R= 20.6. The survey is 97\% complete for R $\leq$ 20.3 and it is 95\% complete for R $\leq$20.6. In the interval 20.3 $< $R$<20.6$ the survey is 89\% complete. The median redshift of the survey for R$\leq$ 20.6 is $z_{med, 20.6} = 0.30$ (see Table \ref{tbl:props}).

The two panels of Figure \ref{Fcomplete} show the two-dimensional completeness of the survey in 12$^{\prime} \times 12^{\prime}$  pixels. Yellow dots indicate galaxies without a redshift measurement. In the central 8$\times$8 pixel region for the survey with R$\leq 20.3$ (left panel), every pixel is more than 92\% complete. Even on the edges, all pixels except  the northwest and southeast corners are
more than 90\% complete. For the survey with 20.3$< {\rm R} \leq 20.6$ (right panel), there are 9 pixels within the central 8$\times 8$ pixel region that are $<90$\% complete, but all of these pixels are more than 80\% complete. For this fainter sample, the edges and corners contain pixels with a completeness $\lesssim 50$\%.  

\begin{figure}
\centerline{\includegraphics[width=7.0in]{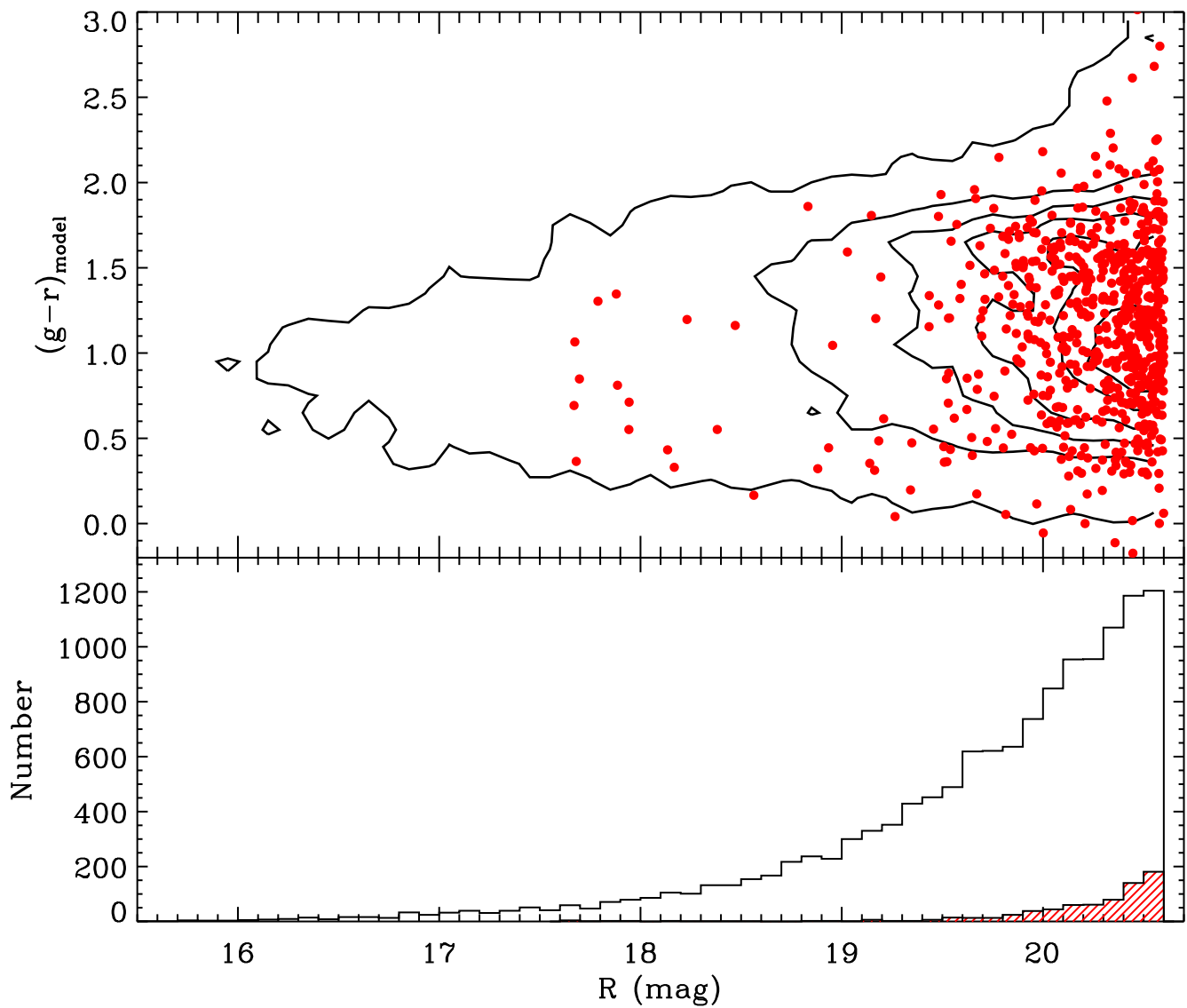}}
\vskip 5ex
\caption{Color-magnitude diagram for the 703 objects without a redshift (upper panel) in the complete survey region. Contours indicate the relative density of objects with a redshift; the absence of a slope as a function of R suggests that there is little obvious color bias in these objects. The lower panel shows the number of objects with redshifts in the SHELS survey (open histogram) and the number of unobserved galaxy candidates (red hashed histogram) as a function of DLS R magnitude.
\label{Fcmr}}
\end{figure}

Figure \ref{Fcmr} shows the distribution of the 703 galaxy candidates in the complete F2 survey region without a measured redshift. We show the observed $(g-r)_{model}$ color as a function of the DLS R-band magnitude for these objects.  The color range for these objects is broad and comparable  with the observed range for the entire sample. There is no significant trend in color with magnitude.

\begin{figure}
\centerline{\includegraphics[width=7.0in]{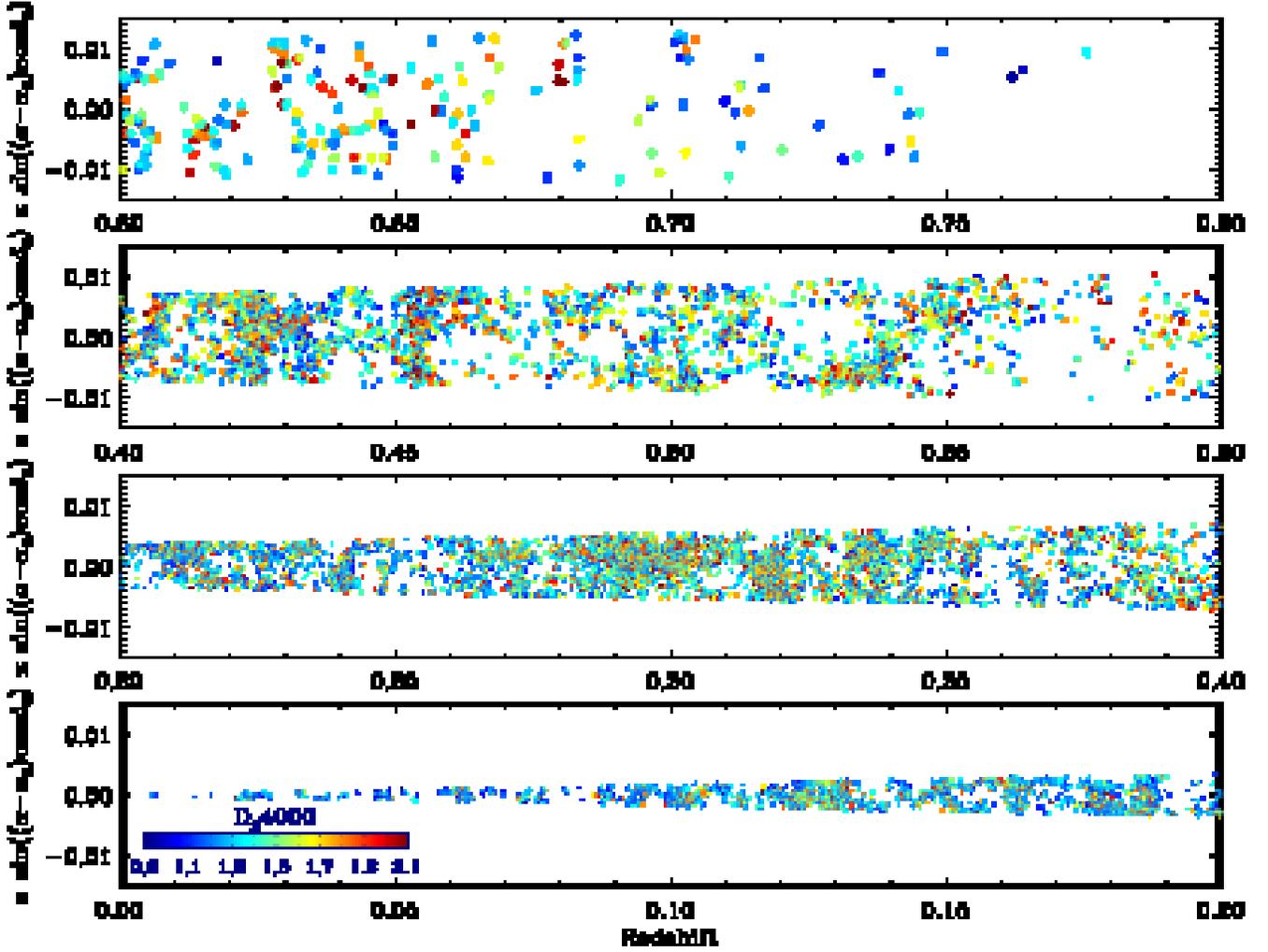}}
\vskip 5ex
\caption{Cone diagram for the R$\leq$20.6 SHELS F2 survey projected on the R.A.$_{2000}$ direction. The color coding indicated the value of D$_n$4000. In the low density regions, galaxies with D$_n$4000 $\lesssim$ 1.5 predominate as expected. The online journal includes a video display of the data. The color-coding of the video is in broader bins: D$_n$4000 $< 1.3$ (blue), 1.3 $\leq$ D$_n$4000 $< 1.7$ (green), and D$_n$4000 $\geq$ 1.7 (red). 
\label{raFcone}}
\end{figure}

Figure \ref{raFcone} shows a cone diagram projected along the R.A.$_{2000}$ direction. The points representing individual galaxies are color-coded with the value of D$_n$4000. The trend toward galaxies with younger stellar populations in lower density regions is evident in the bluer color of the points. An accompanying movie  makes the 3D structure apparent. In the movie we choose broader bins in D$_n$4000. This choice enhances the fingers corresponding to clusters of galaxies.
The most impressive set of clusters lies at $z \sim 0.3$ and the movie zooms in on this region. Zooming in at greater redshift where the survey subtands a larger spatial scale, the movie highlights the emptiness of the large voids along with the thin web-like structures that separate them.

\subsection {Galaxy Colors}
\label{Scolor}
\begin{figure}
\centerline{\includegraphics[width=7.0in]{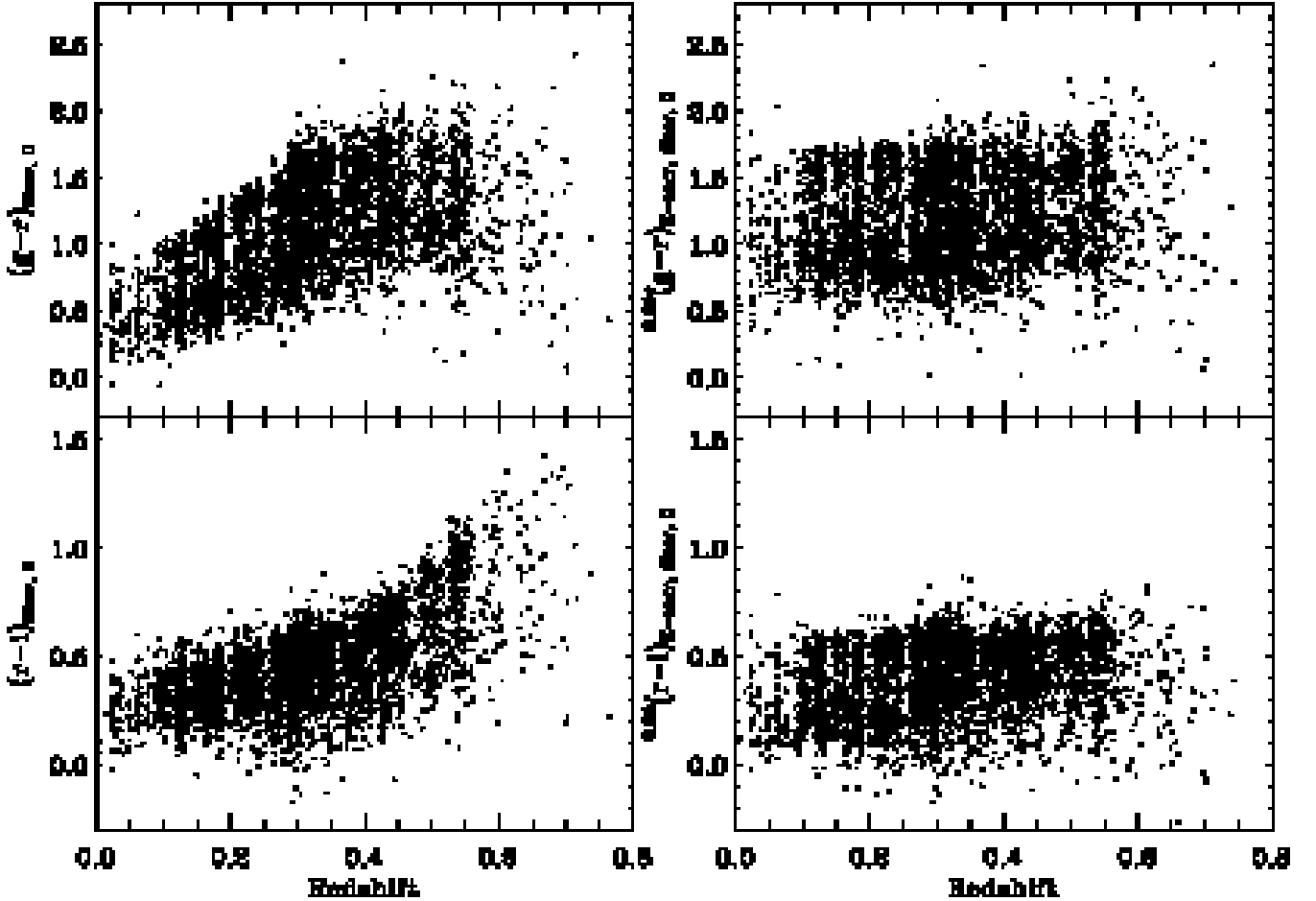}}
\vskip 5ex
\caption{Observed $(g-r)_{fiber,0}$ and $(r-i)_{fiber,0}$ SDSS extinction-corrected colors for the SHELS sample (left panels). (Right) K-corrected colors for SHELS galaxies shifted to the approximate median survey redshift, $ z = 0.35$. We display only 30\% of the data for clarity. The narrowing color distribution at the largest redshifts
occurs because the magnitude limited sample contains increasingly luminous galaxies that are generally somewhat redder.
\label{Fzcol}}
\end{figure}

Figure \ref{Fzcol} shows the observed colors $(g-r)$ and $(r-i)$ of the survey galaxies as a function of redshift. The expected impact of the redshift is obvious. The right-hand panel of Figure \ref{Fzcol} shows the rest frame color of the survey galaxies K-corrected to  $ z = 0.35$, somewhat greater than the median survey redshift. We compute the K-correction using the SDSS $ugriz$ photometry and the kcorrect code of Blanton \& Roweis (2007). The choice
of $z = 0.35$ avoids large relative k-corrections (Blanton et al. 2003a, b) over the range $z \gtrsim 0.2$.

It is interesting that the rest frame color range for the survey galaxies is nearly constant from $z \sim 0.1$ to $z \sim 0.35-0.4$ and then narrows somewhat at greater redshift. The increasing absence of bluer objects at large redshift results from the nature of a magnitude limited survey. At large redshift we observe only the intrinsically most luminous galaxies and they tend to be redder. At $z \lesssim 0.1$ the intrinsic density of the F2 field is low as a result of the DLS selection against nearby clusters and the galaxy population is correspondingly blue. The presence of rich clusters of galaxies at greater redshift is also evident in the color distribution; for example, the dense knot of red galaxies at $z \sim 0.3$ results from the A781 complex.

\section {D$_n$4000 Distributions as Evolutionary Markers}
\label{Smarkers} 
The available photometric data along with the SHELS spectroscopy for the F2 field
provide a uniform, well-defined sampling of galaxy properties in the redshift range 
$0.1 \lesssim z \lesssim 0.6$. In Sections \ref{Sbreak} and \ref{Smass} we display indicative quantities derived from the data as a brief guide to some possible future applications of this dataset. 

We focus here on the continuum feature D$_n$4000 as a function of redshift and stellar mass.
D$_n$4000 is an observable dependent on the  age and metallicity of a stellar population (e.g. Balogh et al. 1999; Kauffmann et al. 2003). In general, objects with large D$_n$4000 have a dominant older stellar population; galaxies with small D$_n$4000 have a predominantly younger population. Using the SDSS main sample of galaxies with 14.5 $< r <$ 17.77 and median redshift $z \sim 0.09$, Kauffmann et al (2003) demonstrate that the distribution of
D$_n$4000 is bimodal reflecting the populations of quiescent and star-forming galaxies.
They also demonstrate that galaxy mass-to-light ratios are primarily a sequence in 
D$_n$4000, but that at fixed D$_n$4000, galaxies with strong H$\delta$ absorption have smaller mass-to-light ratios (these galaxies have undergone a burst of star formation in the last Gyr). In Sections \ref{Sbreak} and \ref{Smass} we show the D$_n$4000 distribution for the SHELS F2 data. We display the observed dependence of the distribution of D$_n$4000 on redshift and stellar mass in the magnitude limited sample.

Many other investigations explore galaxy evolution in the redshift range covered by SHELS F2. To highlight issues where the SHELS data might 
contribute, we briefly review a few examples of previous investigations that use the indicator D$_n$4000. 
Bundy et al. (2006) use DEEP2 data (Davis et al. 2003; Newman et al. 2013) to demonstrate the downsizing (Cowie et al. 1996) of star formation activity in galaxies over the redshift range $0.4 < z < 0.7$. They cleanly demonstrate the downward evolution in the transition mass where galaxies are quenched. Along the way to this conclusion, they emphasize the importance of spectroscopy in the study of galaxy evolution both in the determination of galaxy properties and in the measurement of the densities of their environments. They show that the fractional abundances of quiescent (red) and star-forming (blue) galaxies are relatively insensitive to selection effects in their survey; the quiescent fraction rises with decreasing redshift in each mass bin and the star-forming fraction correspondingly falls. 

The full DEEP2 sample is color-selected to identify galaxies at $z > 0.75$. In the 
extended Groth strip covering 0.6 deg$^2$, the sample is simply magnitude limited 
(R$_{AB} < 24.1$; Newman et al. 2013). Bundy et al. (2006) base their analysis on 943 galaxies on the range 0.4 $< z < 0.7$ overlapping our survey; the SHELS F2 field contains 3371 galaxies within this redshift range and within the complete survey region. SHELS F2 thus complements DEEP2 with a large sample of luminous objects covering a larger area within this redshift range.  In Sections \ref{Sbreak} and \ref{Smass} we use 
the distributions of D$_n$4000 to demonstrate that some of the evolutionary signatures in the DEEP2 data are also evident in our data on a scale more finely binned in redshift. 

A number of surveys explore the detailed evolution of early type galaxies
over the redshift range covered by SHELS (e.g. Roseboom et al. 2006; Moresco et al. 2010; 2013). Roseboom et al. (2006) extract a large sample of $\sim 6000$ luminous red galaxies (LRGs) from the 2dF-SDSS survey. They use D$_n$4000, [OII], and H$\delta$ to explore the star formation histories of the galaxies in their sample. They demonstrate that the fraction of galaxies that have experienced a burst of star formation in the last Gyr increases with redshift over the range $0.45 < z < 0.8$. The color cuts defining the survey restrict the range of star formation histories. The SHELS F2 survey can be used to calibrate color selected surveys because it has the advantage that there are no cuts other than the magnitude limit. 

In analyzing the extensive zCOSMOS samples that cover the 2 deg$^2$ COSMOS field (Scoville et al. 2007; Lilly et al. 2009),
Moresco et al. (2010) extract $\sim 1000$ early-type galaxies (ETGs) in the redshift range $0.45 < z < 1$ from the 10K COSMOS redshift survey covering the magnitude range 
15$ < {\rm I} < 22.5$. They use D$_n$4000 as a discriminant of ETG evolution as a function of stellar mass. They find that in a given mass range D$_n$4000 decreases with increasing redshift. They also use their data to demonstrate that stellar mass is more important than environment in determining the evolution of ETGs. In the redshift range overlapping SHELS F2, zCOSMOS includes lower luminosity, lower mass galaxies, but the survey is sparse and covers only half of the area of SHELS F2.

Moresco et al. (2010) emphasize the importance of spectroscopic data for defining galaxy environments. The  typical $\sim 50$ km s$^{-1}$ error in the Hectospec absorption-line redshifts are smaller than the $\sim$100 km s$^{-1}$ errors for the zCOSMOS redshifts. This difference can be important when measuring the velocity dispersion in group environments and within the web-like structures that define the galaxy distribution.
 
More recently Moresco et al. (2013) have used the zCOSMOS 20K redshift survey to revisit the evolution of ETGs. Interestingly, they find a relationship between the apparent evolution of the sample and the basis for the sample definition (e.g. spectral features, color, and/or morphology). Their samples imply strong evolution of galaxies with stellar masses $\lesssim$10$^{11}$M$_\odot$ and slow or no evolution for more massive galaxies. The most massive galaxies are the least sensitive to the sample definition. 
Their Figure 6 shows stacked spectra binned in mass for ETGs defined in different ways. 
These spectra for $z < 0.5$ can be compared with our example stacked spectra (see Section \ref{Smass}). An advantage of the SHELS F2 data in the mutually explored redshift range is that our spectra include [OII] at low redshift. 

In the following two subsections \ref{Sbreak} and \ref{Smass}, we show that many
of the evolutionary effects uncovered previously are evident in the SHELS F2 data. Thus they offer a platform for enhancing and/or expanding  the existing studies. Because the SHELS F2 sample is strictly magnitude limited, it may serve as a calibrator for subtle  biases in sparse and color-selected surveys to similar depth.

\subsection {D$_n$4000}
\label{Sbreak}
D$_n$4000 is an advantageous measure of galaxy spectra because it is  insensitive to reddening. Furthermore, unlike galaxy color, D$_n$4000 requires no K-correction. Fabricant et al. (2008) consider the sensitivity of D$_n$4000 to aperture effects by comparing 358 
SDSS and Hectospec spectra for the same objects. The galaxies have a median redshift $z \sim 0.1$. Thus the SDSS 3$^{\prime\prime}$ aperture subtends a diameter of 5.5 kpc and includes
$\gtrsim$ 20\% of the light even for luminous galaxies (see Figures 6 and 7 of Kewley et al. 2005). Although the 1.5$^{\prime\prime}$ Hectospec fiber subtends a diameter of only
2.8 kpc, the median ratio of D$_n$4000 values derived from SDSS and Hectospec spectra is 1.0. In other words, even for small apertures there is no apparent bias.  We assume in the following discussion 
that the values of continuum index D$_n$4000 are representative, but we caution that there may be subtle aperture effects. Tests for these effects require larger aperture spectroscopy and/or detailed analysis of surface photometry both of which are beyond the scope of this paper.

Several groups including Mignoli et al. (2005), Vergani et al. (2008), and Freedman Woods et al. (2010) have used D$_n$4000 to segregate galaxies dominated by an old population  from those with a young population. Freedman Woods et al. (2010) show that D$_n$4000 is also well-correlated with the presence/absence of emission lines; the emission-line fraction declines steeply from D$_n$4000 = 1.3 to D$_n$4000 = 1.5. 

\begin{figure}
\centerline{\includegraphics[width=7.0in]{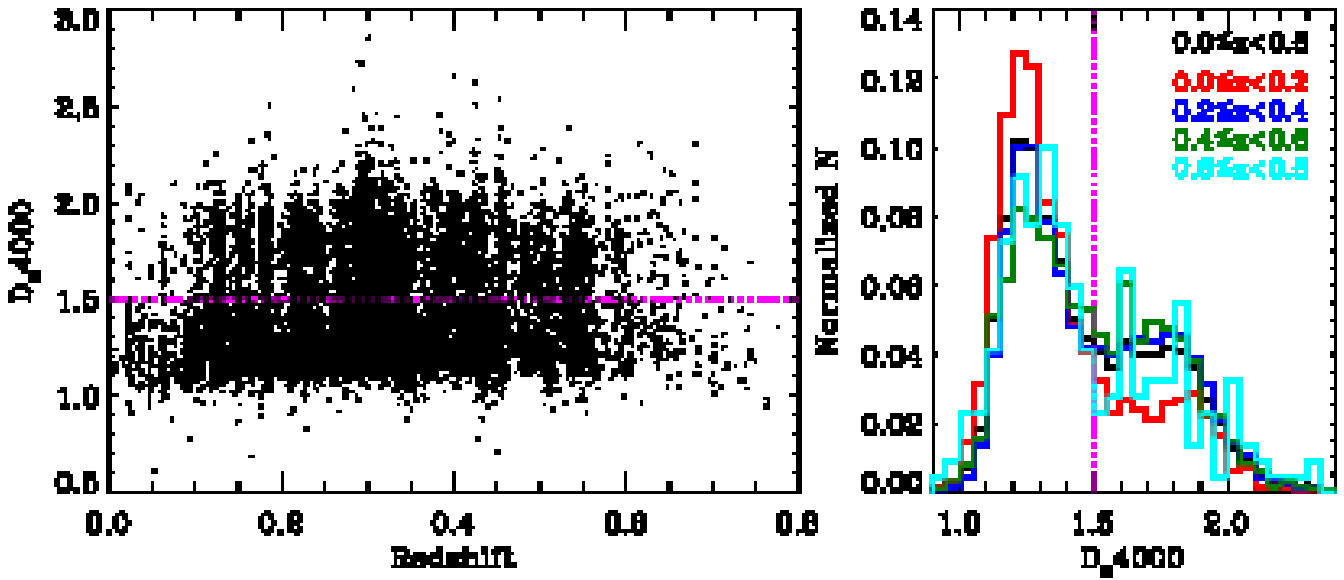}}
\vskip 5ex
\caption{D$_n$4000 as a function of redshift (left) and the normalized distribution of
D$_n$4000 in redshift bins (right). The normalized distributions reflect the complex interplay of 
the magnitude limited sample and the large-scale structure in the survey.  
\label{Fdn4000z}}
\end{figure}

Figure \ref{Fdn4000z} shows the spectral indicator D$_n$4000 as a function of redshift. The fiducial value D$_n$4000 = 1.5 provides an indicative separation between absorption and emission-line systems. At redshift
$ z \lesssim 0.2 $, D$_n$4000 values cluster strongly below the fiducial value of 1.5. This behavior again reflects the selection against nearby clusters in the F2 field. The low density foreground is preferentially populated by star-forming galaxies. Values of D$_n$4000 outside the well populated range (and particularly the unphysical D$_n$4000$ \lesssim 0.8$) result from lower signal-to-noise-spectra.

The right-hand panel of Figure \ref{Fdn4000z} shows normalized histograms of the D$_n$4000 distribution for the entire survey (black) and for selected intervals in redshift. The line at D$_n$4000 = 1.5 roughly segregates the galaxies dominated by younger (D$_n$4000 $ < 1.5)$ and older stellar populations (D$_n$4000 $ > 1.5$). In the interval 0.0 $ < z < 0.2$ there is a relative excess of galaxies with recent star formation, a reflection of the DLS selection for a low-density foreground. In general these histograms demonstrate the complex interplay between the magnitude limit and large-scale structure in the survey. In Section 
\ref{Smass} we show the more easily interpreted distributions of D$_n$4000 segregated by stellar mass and redshift.

\begin{figure}
\centerline{\includegraphics[width=6.0in]{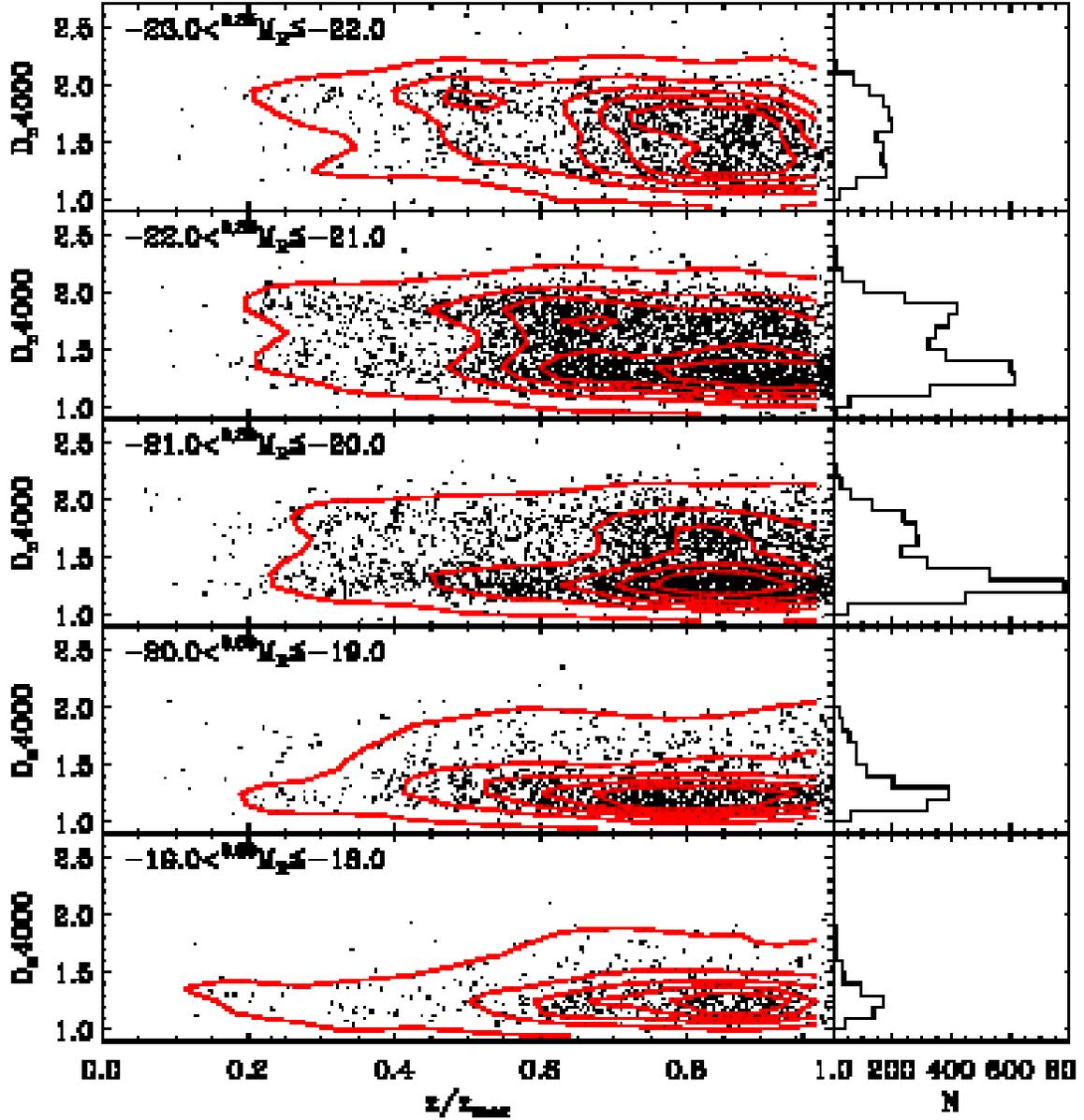}}
\vskip 5ex
\caption{For each K-corrected magnitude interval, we plot (left-hand panels) the value of D$_n$4000 as a function of the ratio $z/z_{max}$ where $z$ is the redshift of the galaxy and $z_{max}$ is the redshift where the galaxy would fall out of the magnitude limited sample.  The shape of the contours in each interval reflects the proportion of galaxies with young/old stellar populations. The typical population is younger for less luminous galaxies as expected. The right-hand panels show histograms of D$_n$4000 in the magnitude interval.
\label{magDn4000}}
\end{figure}

Distributions of D$_n$4000 binned in absolute magnitude show some of the salient evolutionary effects that have been observed in other samples (e.g. Mignoli et al. 2005; Noeske et al. 2007; Vergani et al. 2008; Moustakas et al. 2013). Histograms in the right-hand panels of Figure \ref{magDn4000} show the D$_n$4000 distributions for intervals in absolute magnitude. 
Not surprisingly the fraction of galaxies with D$_n$4000$< 1.5$ increases with decreasing luminosity. The left-hand panels of Figure \ref{magDn4000} show the distribution of D$_n$4000 in each absolute magnitude interval as a function of a normalized redshift 
$z/z_{max}$. Here $z_{max}$ is the redshift where the galaxy apparent magnitude is too faint for inclusion in the survey (Kauffmann et al. 2003). This approach places all galaxies in the interval at the same relative redshift. Because the range of absolute magnitudes we consider in each panel is narrow, the main effect of the scaling is
to smooth out the impact of large-scale structure. 
Clearly, galaxies at greater $z$ are generally at greater $z/z_{max}$. 

The known evolutionary effects are evident in Figure \ref{magDn4000}. At greater $z/z_{max}$, the fraction of galaxies with small D$_n$4000  increases. The effect is stronger for galaxies in the interval $-22.0 <$$ ^{0.35}{\rm M}_R$$ < -20.0$ than for the most luminous galaxies with $-23.0 < $$^{0.35}{\rm M}_R$$< -22.0$ (here the symbols for the absolute magnitudes indicates that they are shifted to $z =0.35$). This effect is the well-known ``downsizing '' where star formation predominates in less and less luminous galaxies as the universe evolves (e.g. Cowie et al. 1996; Bundy et al. 2006; Noeske et al. 2007; Moustakas et al. 2013). At the lowest luminosities we sample, 
$-20.0 <$$ ^{0.35}{\rm M}_R$$ < -18.0$, there are few galaxies with large D$_n$4000 as expected.

\subsection{Stellar Masses}
\label{Smass}

\begin{figure}
\centerline{\includegraphics[width=6.0in]{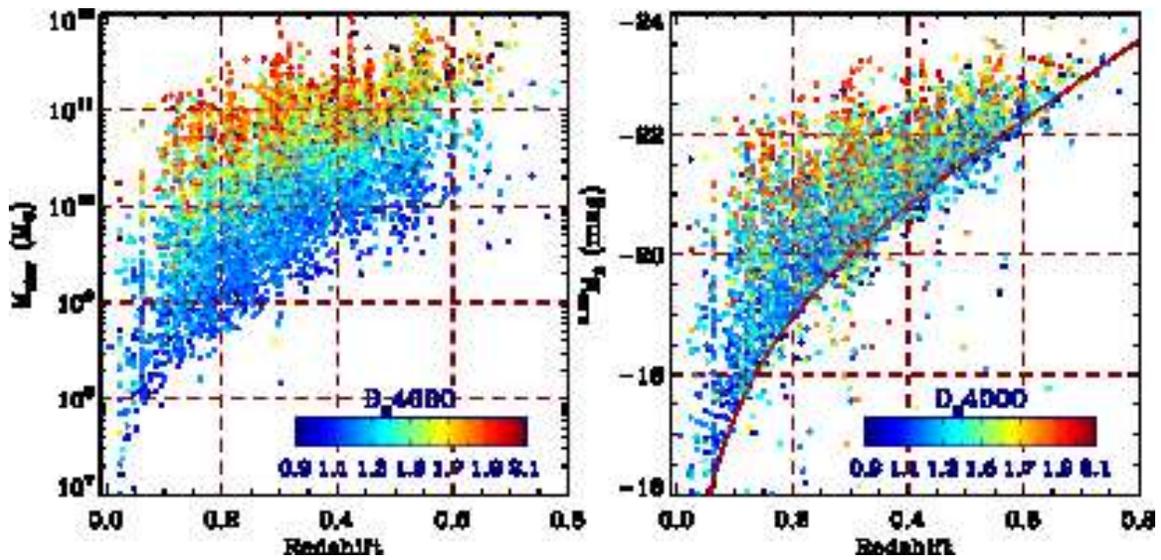}}
\vskip 5ex
\caption{Stellar mass as a function of redshift (left) and K-corrected (to $ z = 0.35$) R-band absolute magnitude as a function of redshift (right). In the both panels galaxies are color-coded by D$_n$4000. We display only 50\% of the data for clarity. The known evolutionary trend that at fixed stellar mass galaxies have younger stellar populations  at greater redshift is evident.  
\label{Fmassz}}
\end{figure}

To gain more insight into the uses of SHELS for exploring issues in galaxy evolution, we show (Figure \ref{Fmassz}) the distribution of stellar masses of SHELS galaxies as a function of redshift (left-hand panel). We color code the galaxies with the value of D$_n$4000. The impact of the limiting survey magnitude is obvious: the mass range narrows with increasing redshift and the galaxies are increasingly massive. 

Stellar mass and D$_n$4000 are not completely independent quantities. However, the stellar mass determination depends on the full $ugriz$ spectral energy distribution. We determine the stellar masses from the photometry alone; we take only the redshift from the spectroscopic data. Thus it is interesting to explore Figure \ref{Fmassz} in some detail. For stellar masses ${\rm M}_{star} \gtrsim 10^{10}$ M$_\odot$, the SHELS survey is deep enough to sample the mass range well to $z \sim$ 0.6. It is qualitatively obvious that in the fixed mass range $10^{10} < {\rm M}_{star} < 10^{11}$M$_\odot$, the  typical value of D$_n$4000 decreases as the redshift increases. The effect is less dramatic for the most massive galaxies with
${\rm M}_{star} > 10^{11}$ M$_\odot$. Similar evolution is apparent in, for example, the PRIMUS survey (Moustakas et al. 2013)

The spectral evolution of the galaxy population is not nearly as evident in the right-hand panel of Figure \ref{Fmassz} where we display the absolute magnitude $^{0.35}{\rm M}_R$
as a function of redshift color-coded by D$_n$4000. In essence the two panels of Figure
\ref{Fmassz} underscore the importance of use of the entire spectral energy distribution 
in elucidating the way galaxies evolve.

\begin{figure}
\centerline{\includegraphics[width=6.0in]{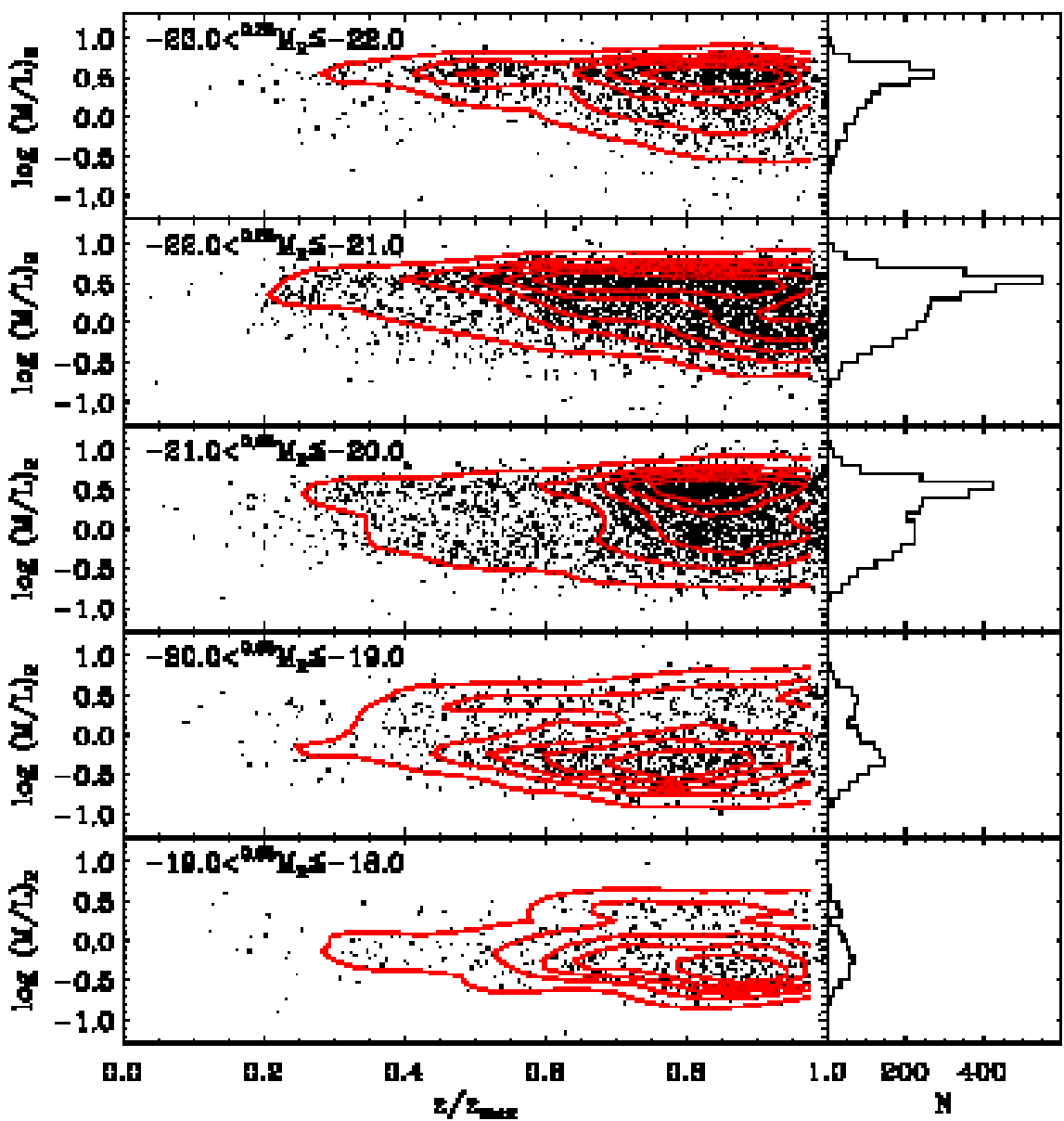}}
\vskip 5ex
\caption{R-band mass-to-light ratio as a function of $z/z_{max}$; $z$ is the galaxy redshift and $z_{max}$ is the redshift where the galaxy becomes fainter than the magnitude limit.
The panels show bins in K-corrected absolute magnitude. Note that the contours have no significant slope with $z/z_{max}$. Note also the peak of the M/L$_R$ distribution moves toward smaller M/L$_{R}$ for less luminous galaxies as expected.   
\label{magAmlratio}}
\end{figure}

We can combine the two panels of Figure \ref{Fmassz} to examine the R-band mass-to-light ratio of galaxies (in solar units) as a function of their absolute magnitude and normalized redshift
$z/z_{max}$. Figure \ref{magAmlratio} shows the result. 
For the most luminous galaxies, $-23.0 < $$ ^{0.35}{\rm M}_R $$ < -21.0$, the mass-to-light ratio distribution peaks near log(M/L)$_R \simeq$ 0.6 with a more substantial tail toward smaller values for less luminous galaxies. For $-21.0 < $$^{0.35}{\rm{M}}_R$$ < -20.0$, there is a secondary peak at log(M/L)$_R \simeq -0.2$ resulting from a larger populations of galaxies with a young stellar population. For galaxies with 
$-20.0 < $$^{0.35}{\rm{M}}_R$$ < -18.0$, galaxies with young stellar populations predominate and the mass-to-light ratios are generally low.

Comparison of the plots in Figures \ref{magDn4000} and \ref{magAmlratio} seems at first glance to present a puzzle. In the magnitude interval
$-22.0 < $$^{0.35}{\rm{M}}_R$$ < -20.0$, the distribution of D$_n$4000 peaks at 1.3 (Figure \ref{magDn4000}) but log(M/L)$_R$ peaks at $\sim 0.6$ (Figure \ref{Fmassz}) typical of a population with 1.5 $< $D$_n$4000 $< 2$. 

\begin{figure}
\centerline{\includegraphics[width=7.0in]{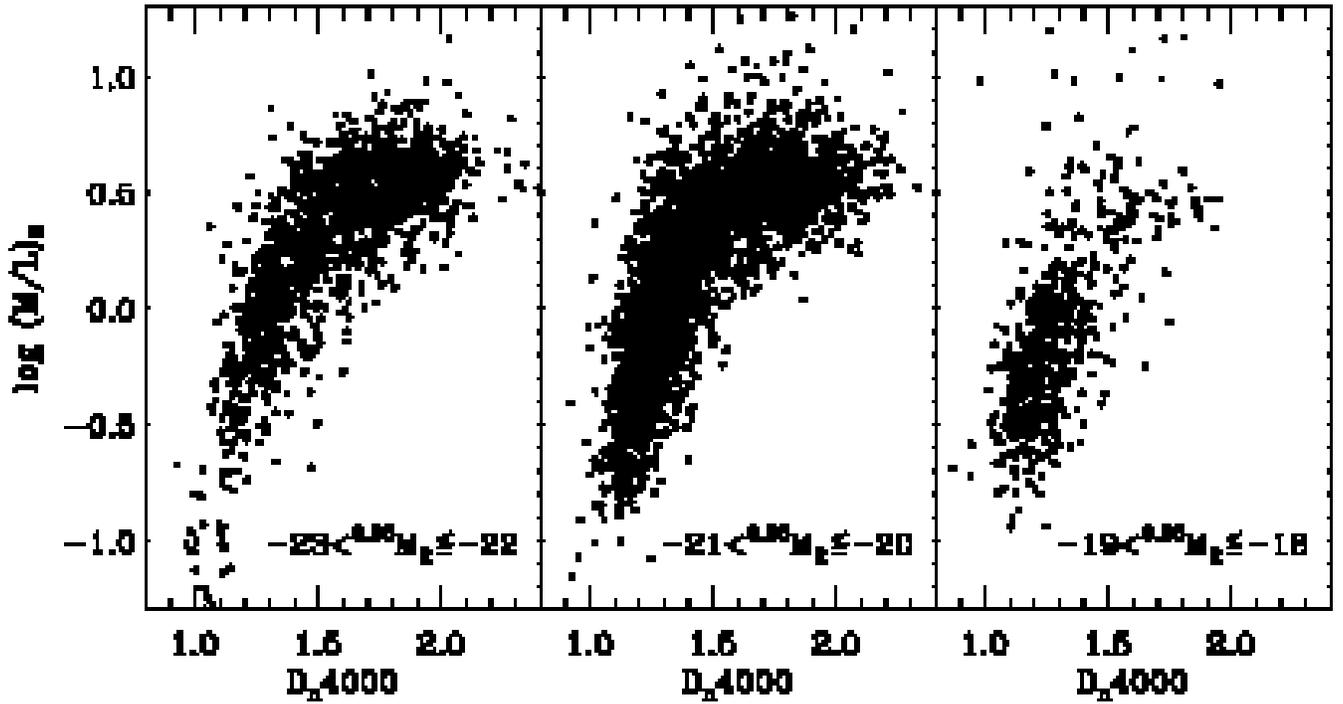}}
\vskip 5ex
\caption{R-band mass-to light ratio M/L$_R$ as a function of D$_n$4000 for three luminosity bins. Note the steep relation for D$_n$4000$\lesssim$ 1.5 (typically star-forming galaxies) and the shallow dependence for
D$_n$4000$ \gtrsim$1.5 (typically quiescent galaxies).
\label{Fpmld4000}}
\end{figure}

The dependence of log(M/L)$_R$ and D$_n$4000 shown in Figure \ref{Fpmld4000} explains the relationship between Figures \ref{magDn4000} 
and \ref{magAmlratio}. For the predominantly star-forming galaxies with D$_n$4000 $\leq$ 1.5,
log(M/L)$_R$ is a steep function of D$_n$4000; thus for these objects a narrow range in D$_n$4000 translates to a broad range in log(M/L)$_R$. In other words, a peak in
the distribution of D$_n$4000 stretches out in the corresponding distribution of
(M/L)$_R$. For  D$_n$4000$ > 1.5$, the corresponding (M/L)$_R$ range is small; here a broad distribution in D$_n$4000 becomes a narrow peak in (M/L)$_R$. The three panels of Figure \ref{Fpmld4000} show that the relationship between D$_n$4000 is similar throughout the range of absolute magnitudes.

\begin{figure}
\centerline{\includegraphics[width=7.0in]{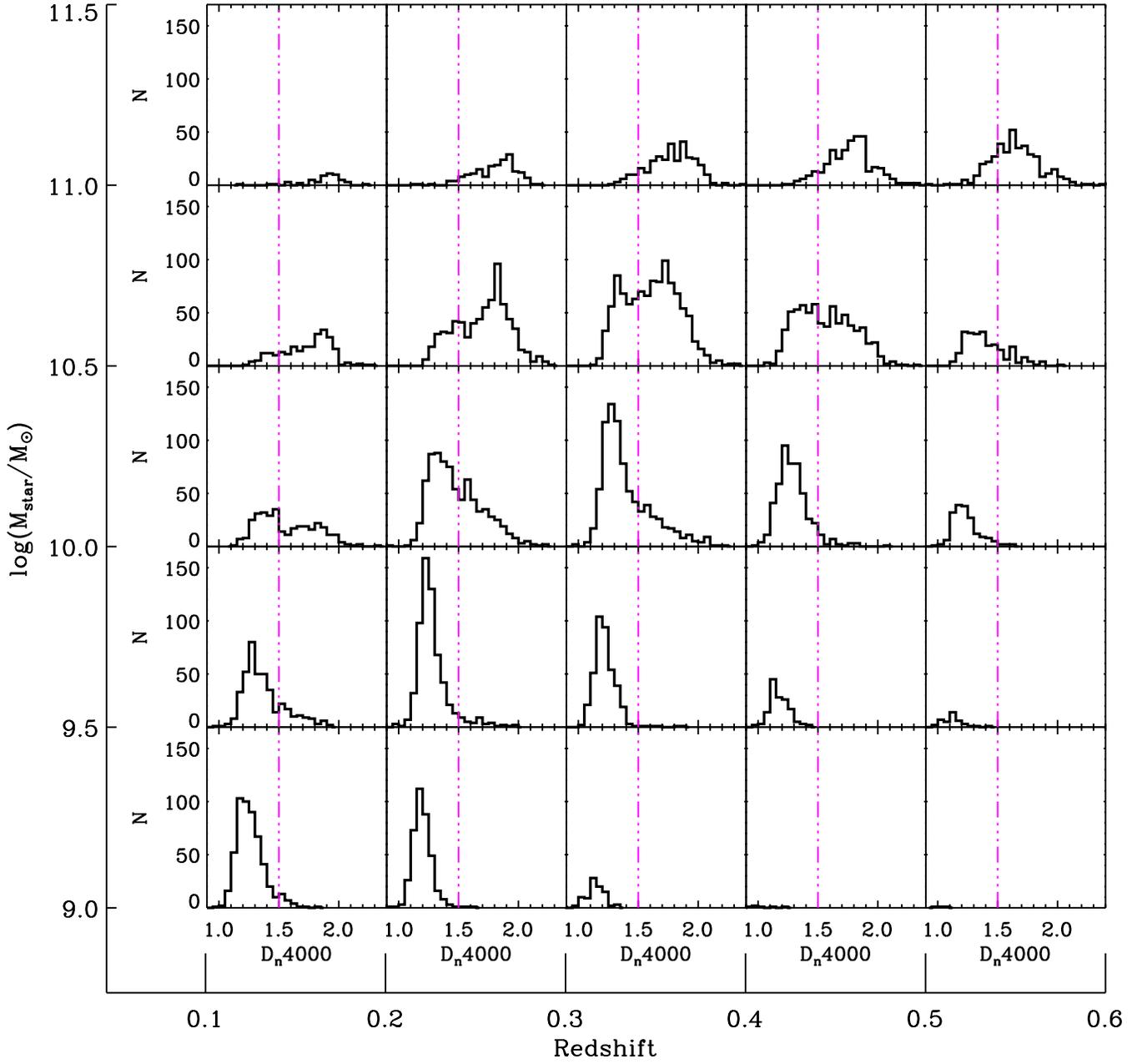}}
\vskip 5ex
\caption{Histograms of D$_n$4000 in bins of stellar mass and redshift. At fixed stellar mass, the expect evolutionary effects appear; the fraction of low D$_n$4000 (probable star-forming galaxies) increases with redshift at fixed stellar mass. 
\label{subd4000}}
\end{figure}

Figure \ref{subd4000} shows another representation of the data displayed in the left-hand panel of Figure \ref{Fmassz}. The histograms of D$_n$4000 as a function of stellar mass and
redshift reveal more subtle effects. For the most massive objects
(10$^{11} < {\rm M_{star}/M_\odot} < 10^{11.5}$), the distribution develops a relatively large tail with D$_n$4000 $\leq 1.5$ at large redshift. Even in this mass bin, the survey is deep enough to provide suggestive evidence of the expected spectral evolution with redshift in the sense that a larger fraction of these galaxies harbor a young stellar population at larger $z$. This effect is much more pronounced for
galaxies in the stellar mass range 10$^{10} < {\rm M_{star}/M_\odot} < 10^{11}$ where most of the objects in the survey lie. In the bins of lowest stellar mass, the galaxies are predominantly star-forming.

\begin{figure}
\centerline{\includegraphics[width=7.0in]{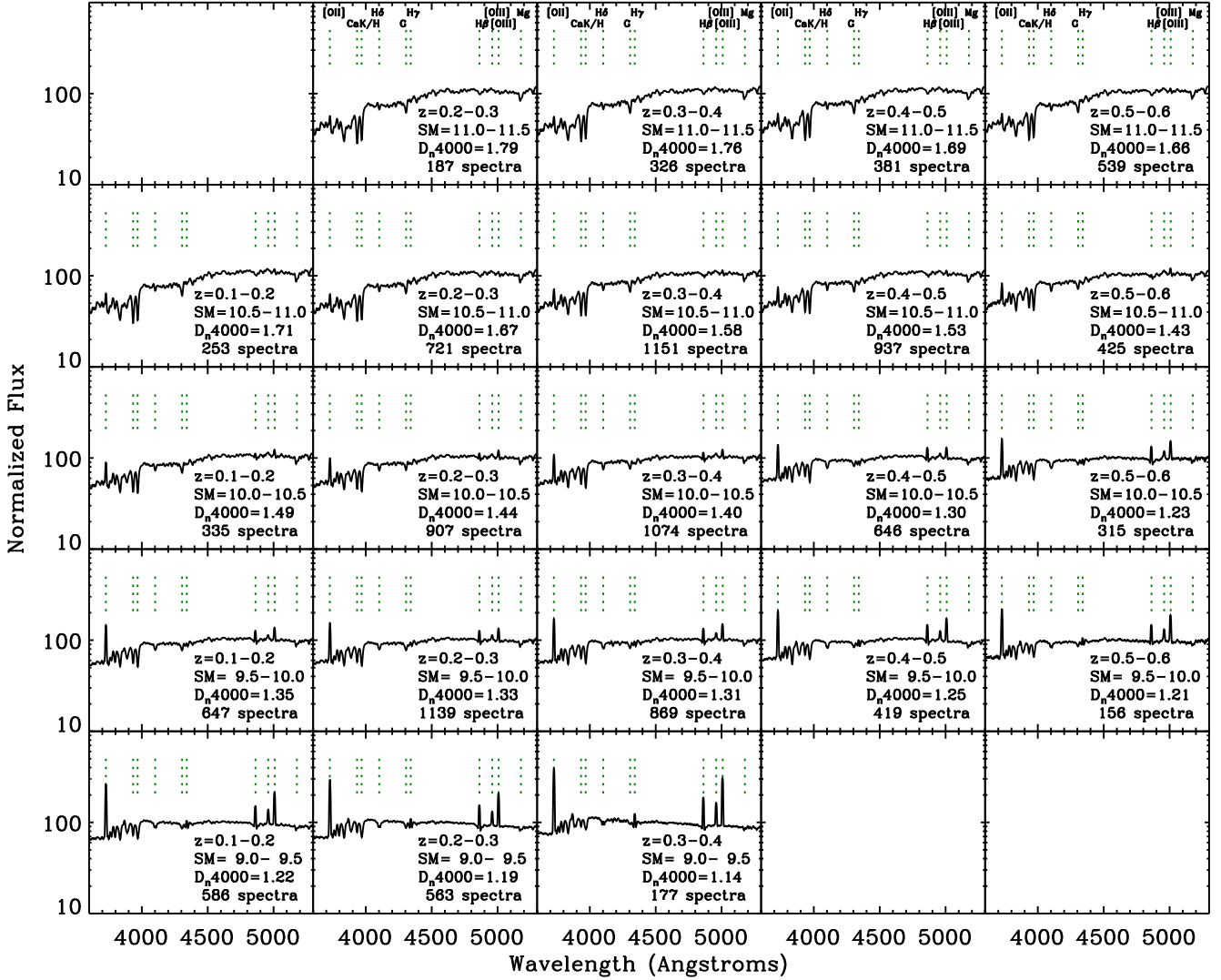}}
\vskip 5ex
\caption{Summed spectra in the same bins as Figure \ref{subd4000}. The caption specifies the redshift range (z), the stellar mass range {SM}, the D$_n$4000 of the summed spectrum, and the number of spectra summed in each case. Dashed vertical lines identify several spectral features. Note the increase in D$_n$4000 with decreasing redshift, the appearance of Balmer absorption lines and emission lines with decreasing stellar mass. These spectra serve as a broad brush guide to the galaxy population in the sample.
\label{sumspec1}}
\end{figure}

Figure \ref{sumspec1} shows the SHELS F2 rest-frame spectra summed in bins of 0.1 in $z$ and 0.5 dex in stellar mass corresponding to the histograms in Figure \ref{subd4000}. We show only the rest frame wavelength range 3600 -5300 \rm \AA\  that we sample throughout the redshift range of the survey. The summed spectra include galaxies with 0.1 $< z < $0.6 and 10${^9}< {\rm M_{star}/M_\odot} < 10^{11.5}$.
The input spectra are flux corrected to units of relative F$_{\lambda}$ (Fabricant et al. 2013), preserving the relative detected fluxes. We reject spectra with with D$_n$4000 $< 0.8$ or D$_n$4000 $> 3.0$.  In Figure \ref{sumspec1} we normalize the final fluxes to a value of 100 at 4500 \rm  \AA.

The summed spectra show once again the increasing impact of star formation at higher z and lower stellar mass. In each panel, labeled by the redshift and stellar mass interval, we also list the value of D$_n$4000 for the summed spectrum. The behavior of the emission lines is also notable, particularly the clear trends of [OII]$\lambda$3727
with D$_n$4000. It is interesting that even for the most massive galaxies in the sample, there is very weak [OII]$\lambda$3727 suggesting that essentially all galaxies have some residual star formation as indicated by studies at near ultraviolet and infra-red wavelengths (Yi et al. 2005; Hwang et al. 2012a; Ko et al. 2013). For the entire sample of summed spectra, the strength of  [OII]$\lambda$3727 is visibly anti-correlated with D$_n$4000. Balmer absorption lines are increasingly evident at larger redshift and for less massive objects.

These summed spectra provide a fiducial guide to the SHELS survey. Table
\ref{tab-fig14} lists the normalized flux as a function of rest-frame wavelength for all of the spectra; these representative spectra may be a useful test set for population synthesis models and perhaps for instructional purposes.

\section{The Mass-Metallicity Relation Through Summed Spectra for $0.2 < z < 0.38$}
\label{SMZrel}

The mass-metallicity (MZ) relation provides another dimension for exploring galaxy evolution. In contrast with the discussion of D$_n$4000, the metallicities we report here use only star-forming galaxies in the redshift range 0.2 $< z <0.38$ where we have access to H$\alpha$.
Zahid et al. (2013, 2014) place the SHELS MZ relation in the context of other surveys. The data show that the mass-metallicity relation flattens as the universe evolves. They also show that the gas-phase metallicity saturates; the saturation level is independent of redshift. 
Zahid et al. (2013) use the  SHELS data for the F2 field as a partial basis for determination of the MZ relation at 0.2$< z < 0.38$. Two aspects of the
Hectospec spectroscopy limit the redshift range. For $ z \lesssim 0.2$ the fibers typically include less than 20\% of the light and the metallicity may be biased. At $z > 0.38$ 
H$\alpha$ is no longer within the Hectospec bandpass and thus the normal BPT diagram cannot
be used to discriminate against AGN. Table \ref{tab-samp} includes metallicities for the individual galaxies in F2 used in the Zahid et al. (2013) analysis. 

Zahid et al. (2013) base their MZ relation on the analysis of individual spectra from both the F1 and F2 fields of the SHELS survey; the F1 field contibutes 40\% of the
spectra in that study.  Here, to investigate the robustness of the 
result, we construct summed spectra in mass bins for the F2 data alone. We then measure a 
metallicity for each summed spectrum and ask whether the results track the relation in 
Zahid et al. (2013). 

\begin{figure}
\centerline{\includegraphics[width=7.0in]{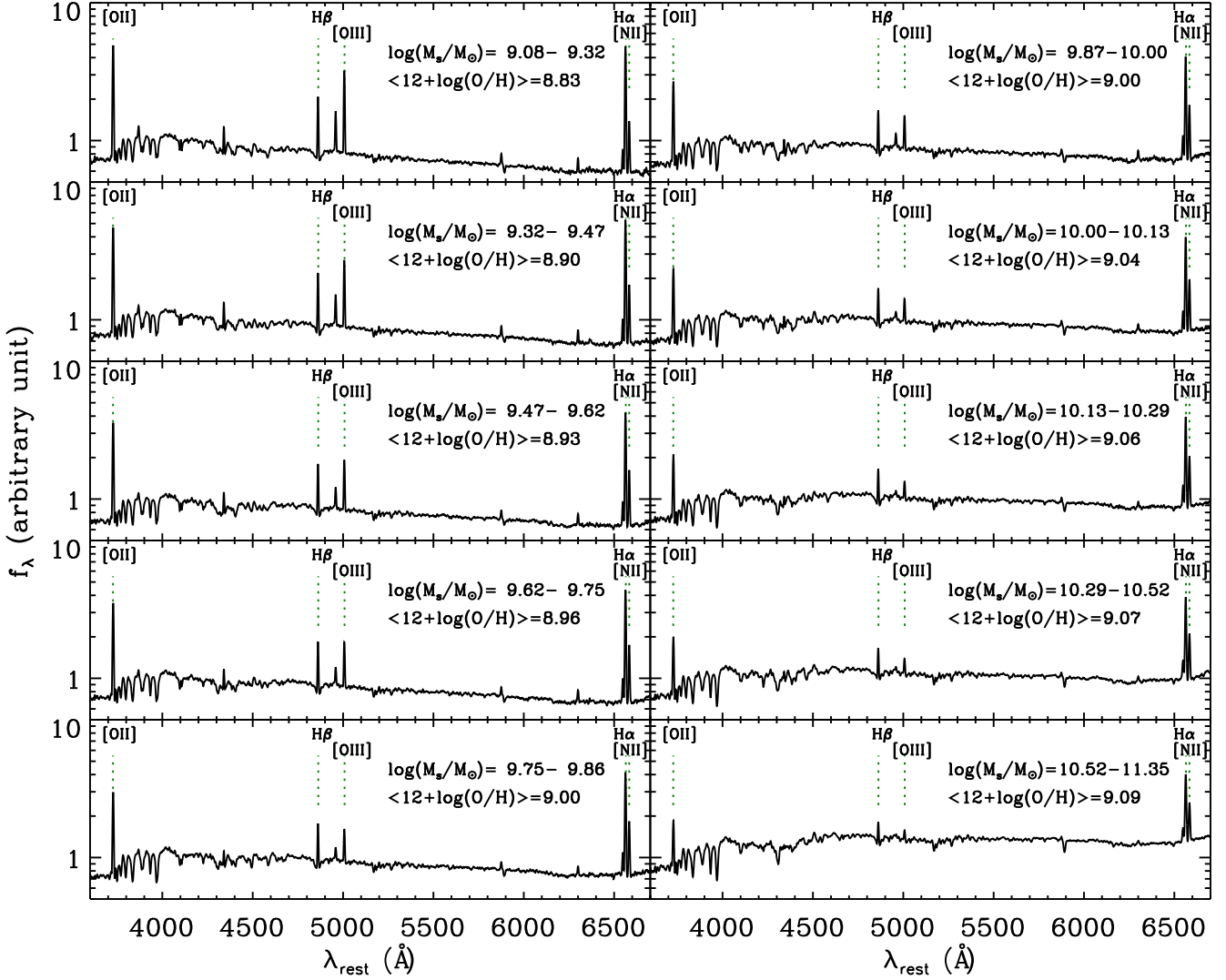}}
\vskip 5ex
\caption{Summed spectra for star-forming emission-line galaxies with 0.2$< z <$0.38 
binned in stellar mass. The legend indicates the mass range and the metallicity derived
by applying the KK04 calibration to the summed spectrum. 
\label{Fspmetal}}
\end{figure}

To construct the summed spectra in Figure \ref{Fspmetal} we apply the selection criteria described in Section \ref{Sspecprops}. In addition we remove 58 objects where the spectrum is contaminated by poorly subtracted night sky lines. The final sample contains 2131 spectra.

To construct the summed spectra, we first sort the spectra into equally populated bins of stellar mass.  Each mass bin (the limits are indicated in the Figure legend) is based on a stack of 213 galaxies. 
We then linearly interpolate the spectra and observational uncertainties to a common rest-frame wavelength vector based on the observed redshift. The interpolated rest-frame wavelength vector has a 1.5 Angstrom resolution and covers the observed wavelength range
3500-9100 \AA. At each resolution element, the stacked spectrum is the average interpolated flux of all spectra in the bin.

The spectra  in Figure \ref{Fspmetal} are ordered from top to bottom and from left to right beginning with the lowest metallicity at the upper left.  The legend gives the stellar mass range or each bin and the KK04 metallicity. Changes in the spectra are obvious by visual inspection. The strength of [OII]$\lambda$3727 decreases and the strength of [NII] relative to H$\alpha$ increases as the metallicity increases. There are corresponding decreases in the strength of
[OIII] and H$\beta$. Table \ref{tab-fig15} lists the normalized flux as a function of rest-frame wavelength for these spectra.

\begin{figure}
\centerline{\includegraphics[width=7.0in]{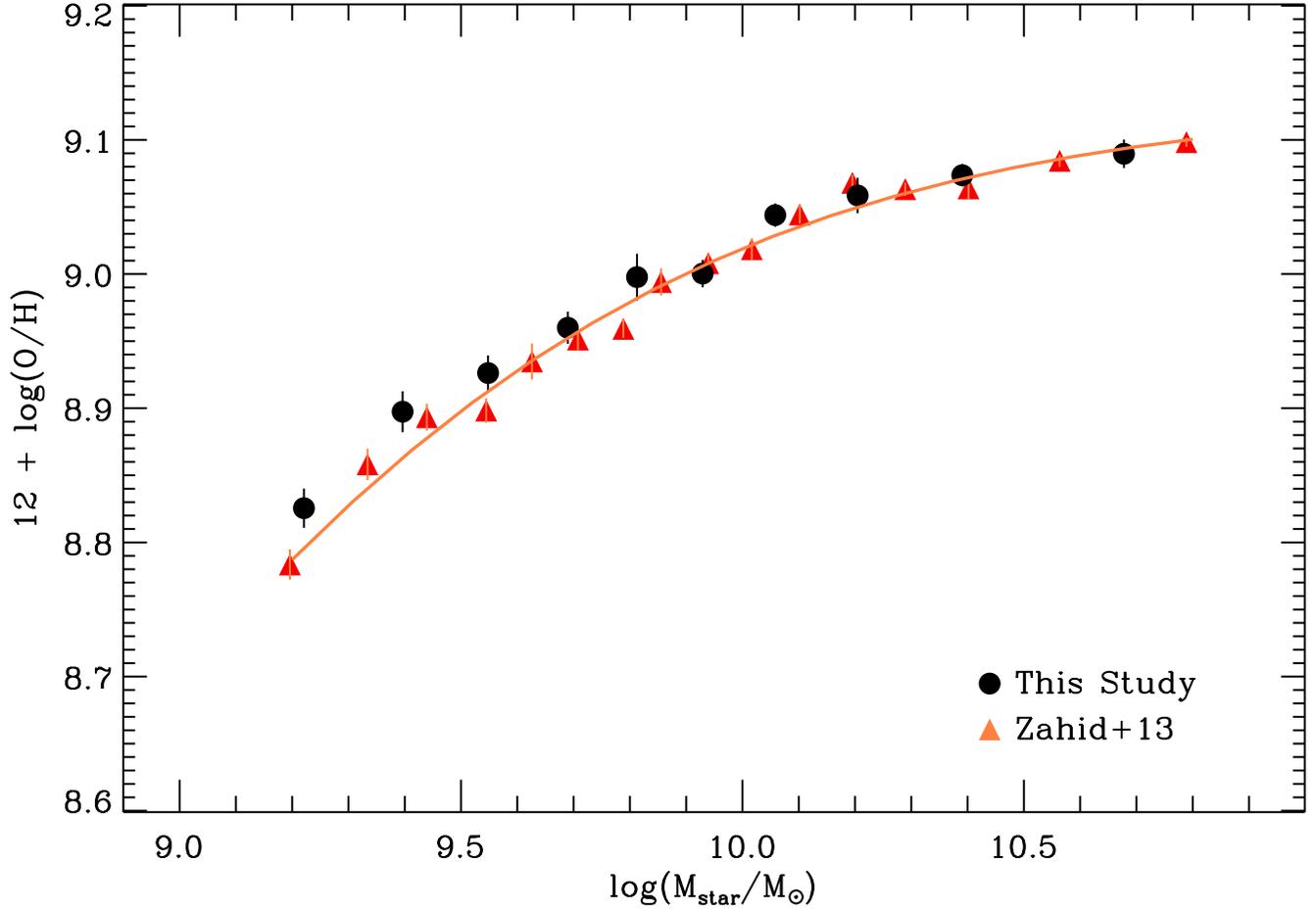}}
\vskip 5ex
\caption{Mass-metallicity relations from Zahid et al. (2013) (red triangles) and from the summed spectra in Figure \ref{Fspmetal} (filled circles).
\label{Fmzrel}}
\end{figure}

Figure \ref{Fmzrel} shows the MZ relation computed from the spectra in Figure \ref{Fspmetal}
compared with the results of Zahid et al (2013). The two estimates of the relation are not independent because Zahid et al. (2013) include the F2 data, but  analyzed in a completely different way. The comparison in Figure \ref{Fmzrel} demonstrates the robustness of the MZ relation to sampling and analysis issues. It also demonstrates the utility of the summed spectra here and, by analogy, in 
Figure \ref{sumspec1}.

\section{Conclusion}
\label{Sconc}

The goal of the SHELS project is to use redshift surveys complete to R = 20.6 to explore the galaxy distribution and related spectroscopic properties of galaxies in the redshift range 0.1$ \lesssim z \lesssim 0.6$. The survey covers two widely-separated DLS fields.
The total area of the survey will be $\sim$ 8 deg$^2$. The first complete field, F2 field 
(R.A.$_{2000}$ = 09$^h$19$^m$32.4$^s$ and Decl.$_{2000}$ = +30$^{\circ}$00$^{\prime}$00$^{\prime\prime}$) includes 12,705 galaxies with R$ \leq$ 20.6 in a 3.98 deg$^2$ clean survey region. The redshift survey is 95\% complete to this limit. Currently the F2 survey is the
densest redshift survey to this limit. The median redshift of the survey $z = 0.3$.

We outline the properties of the survey emphasizing the simple, complete magnitude limited selection. This selection makes the survey a useful calibrator for both sparse and color-selected surveys. 

The Hectospec spectra cover the wavelength range 3,700 --- 9,100\AA~ and the typical error in the redshift (normalized by $(1 + z)$) for an individual galaxy is 
$\lesssim 50$ km s$^{-1}$.
The small redshift errors make the survey very well-suited to evaluation of the local velocity dispersion in the full range of environments. The inclusion of [OII]$\lambda$3727 over the full survey range provides a useful probe of the properties of star-forming galaxies. 

For nearly all of the galaxies in the complete survey region
we provide the redshift, the spectroscopic indicator, D$_n$4000, and an estimate of the stellar mass.  In the narrow redshift range 0.2 $< z < $ 0.38, we provide an emission-line based metallicity. The total dataset includes 16,319 redshifts; 16,294 of these
are new redshifts measured with the Hectospec on the MMT. 

We broadly examine the properties of the survey galaxies as a function of D$_n$4000 and stellar mass as a guide to future applications of the survey. Because the survey selection is straightforward and the completeness is very high, the data should provide an interesting testbed for refining the analyses of galaxy properties as a function of redshift and environment. 

We also demonstrate the robustness of our previously published mass-metallicity relation that is based in part on the F2 survey. The individual metallicities provide a complement to measurements in other redshift ranges.

The second field of the SHELS survey, F1 (R.A.$_{2000}$ = 00$^h$53$^m$25.3$^s$ and
Decl.$_{2000}$ = 12$^\circ$33$^{\prime}$55$^{\prime\prime}$), is not yet complete. We plan to provide similar data for F1 soon. It is interesting that the ratio of galaxy counts to the limiting R = 20.6 in F1 versus F2 is $\sim$0.74. In contrast with F2, F1 contains no complex of rich clusters like A781. Take together the F2 and F1 data will provide a window on the impact of cosmic variance in this redshift range.
 
\begin{acknowledgments}
We thank  Perry Berlind and Michael Calkins who masterfully operated the Hectospec on the MMT and Susan Tokarz who efficiently reduced the data. Sean Moran contributed to the construction of the catalogs contained here. We thank Sangwoo Lee for helping us to make the movie. The Smithsonian Institution supports the research of MJG, HSH, DGF and MJK. 

\end{acknowledgments}

{\it Facilities:}\facility {MMT(Hectospec)}

\clearpage

\begin{deluxetable}{rccr}
\tabletypesize{\footnotesize}
\tablewidth{0pc} 
\tablecaption{List of Masked Regions and Radii\tablenotemark{a}
\label{tab-mask}}
\tablehead{
ID & R.A.$_{2000}$ & Decl.$_{2000}$ & Radius \\
   & (deg)         & (deg)          & (arcsec)
}
\startdata
   1 & 138.7028360 &  30.2579880 &    39.1 \\
   2 & 138.7108755 &  29.7457428 &    38.5 \\
   3 & 138.7169266 &  29.6362076 &    53.5 \\
   4 & 138.7175417 &  29.6406803 &    35.0 \\
   5 & 138.7219620 &  30.1846695 &    10.8 \\
   6 & 138.7447071 &  30.8056984 &    50.4 \\
   7 & 138.7479687 &  29.2265320 &    32.9 \\
   8 & 138.7577963 &  29.5055408 &    30.8 \\
   9 & 138.7579107 &  30.9256687 &    32.9 \\
  10 & 138.7674952 &  29.8776798 &    10.8 \\
\enddata
\tablenotetext{a}{This table is available in its entirety in a machine-readable form in the online journal. A portion is shown here for guidance regarding its form and content.}
\end{deluxetable}
\clearpage

\begin{deluxetable}{lccr}
\tabletypesize{\large}
\tablewidth{0pc} 
\setlength{\tabcolsep}{1.0in}
\tablecaption{\large SHELS F2 Redshift Survey Properties 
\label{tbl:props}}
\tablehead{
Parameter & Value }
\startdata
   Survey Area (deg$^2$) & 3.98\\
   N$_{phot, 20.3}$\tablenotemark{a} & 9946\\
   N$_{z, 20.3}$\tablenotemark{b}& 9643 \\ 
   N$_{phot, 20.6}$&13408\\
   N$_{z, 20.6}$& 12705 \\
   $z_{med, 20.6}$&0.30\\
   N$_{z,mask}$\tablenotemark{c}& 672\\
   N$_{z, R > 20.6}$\tablenotemark{d}& 2994\\
\enddata
\tablenotetext{a}{Number of photometric objects
in the complete survey (unmasked) region brighter than the quoted limit.}
\tablenotetext{b}{Number of galaxies with a measured redshift brighter than the specified limit in the complete survey region.} 
\tablenotetext{c}{Number of galaxies within the masked region of any apparent magnitude (photometry is
unreliable in these regions).}  
\tablenotetext{d}{Number of objects with a redshift in the complete survey region but with R$>20.6$.}
\end{deluxetable}

\clearpage

\begin{deluxetable}{cccccccrc}
\rotate
\tabletypesize{\footnotesize}
\tablewidth{0pc} 
\tablecaption{SHELS Redshifts with $R\leq20.6$\tablenotemark{a}
\label{tab-samp}}
\tablehead{
SHELS ID & SDSS ObjID\tablenotemark{b}  & $R$   & $z$ & $z$                     & Mask\tablenotemark{d} & $D_n4000$ & log($M_\star/M_\odot)$ & 12        \\
         &                              & (mag) &     & Source\tablenotemark{c} &               
        &           &                        & +log(O/H)
}
\startdata
138.7003546+30.7364270 & 1237664093976986160 & $18.951\pm0.041$ & $ 0.27464\pm0.00049$ & 1 & 0 &
 1.65 &                       ... &   ... \\
138.7012673+30.6519668 & 1237664093976986057 & $19.974\pm0.003$ & $ 0.39849\pm0.00014$ & 1 & 0 &
 1.86 &   $10.84^{+0.10}_{-0.17}$ &   ... \\
138.7015818+30.4485196 & 1237664668965601390 & $18.189\pm0.019$ & $ 0.12405\pm0.00009$ & 1 & 0 &
 1.32 &   $10.20^{+0.14}_{-0.07}$ &   ... \\
138.7018711+30.8165660 & 1237664669502538244 & $20.256\pm0.004$ & $ 0.39819\pm0.00016$ & 1 & 0 &
 2.02 &   $10.82^{+0.12}_{-0.14}$ &   ... \\
138.7027976+30.5422030 & 1237664668965601456 & $18.651\pm0.002$ & $ 0.11263\pm0.00022$ & 1 & 0 &
  ... &                       ... &   ... \\
138.7056238+30.3584043 & 1237664668965601602 & $19.755\pm0.002$ & $ 0.32837\pm0.00013$ & 1 & 0 &
 1.66 &   $11.00^{+0.06}_{-0.05}$ &   ... \\
138.7056609+29.9143189 & 1237664668428600041 & $20.588\pm0.005$ & $ 0.36637\pm0.00010$ & 1 & 0 &
 1.17 &   $ 9.55^{+0.23}_{-0.19}$ &  9.02 \\
138.7060648+30.6005998 & 1237664093976985829 & $19.028\pm0.002$ & $ 0.32193\pm0.00006$ & 1 & 0 &
 1.18 &   $10.52^{+0.10}_{-0.11}$ &  9.10 \\
138.7061487+29.6766083 & 1237664092903047686 & $20.337\pm0.004$ & $ 0.36659\pm0.00013$ & 1 & 0 &
 1.67 &   $10.27^{+0.18}_{-0.15}$ &   ... \\
138.7061703+30.1344974 & 1237664093439984063 & $19.392\pm0.002$ & $ 0.26388\pm0.00010$ & 1 & 0 &
 1.22 &                       ... &   ... \\
\enddata
\tablenotetext{a}{This table is available in its entirety in a machine-readable form in the online journal. A portion is shown here for guidance regarding its form and content. The full table contains 13325 galaxies with a measured redshift; 13300 are MMT Hectospec redshifts and 25 are from the SDSS. There are 12705 galaxies in the complete survey region (12680 MMT redshfiifts and 25 SDSS redshifts all denoted 0 in column 6)}
\tablenotetext{b}{SDSS ObjID from DR10. If DR10 ObjID is not available, we present SDSS DR7 ObjID that starts with `58'.}
\tablenotetext{c}{(1) This study; (2) SDSS.}
\tablenotetext{d}{(0) Outside masked regions; (1) Inside masked regions.}
\end{deluxetable}

\clearpage

\begin{deluxetable}{ccccccrc}
\tabletypesize{\scriptsize}
\tablewidth{0pc} 
\tablecaption{SHELS Redshifts with $R>20.6$\tablenotemark{a}
\label{tab-sampfa}}
\tablehead{
SHELS ID & SDSS ObjID\tablenotemark{b}  & $R$   & $z$\tablenotemark{c} &  Mask\tablenotemark{d} 
& $D_n4000$ & log($M_\star/M_\odot)$ & 12        \\
         &                              & (mag) &                      &                        
&           &                        & +log(O/H) 
}
\startdata
138.7021866+30.7739149 & 1237664093976986206 & $20.755\pm0.005$ & $ 0.39146\pm0.00008$ & 0 & 1.24 &   $ 9.78^{+0.22}_{-0.19}$ &   ... \\
138.7045840+30.5183923 & 1237664668965601819 & $20.823\pm0.005$ & $ 0.28946\pm0.00011$ & 0 & 1.25 &   $ 9.31^{+0.19}_{-0.15}$ &  8.94 \\
138.7067846+30.2999792 & 1237664093440049607 & $20.891\pm0.006$ & $ 0.50448\pm0.00014$ & 0 & 1.17 &   $ 9.91^{+0.36}_{-0.25}$ &   ... \\
138.7072475+29.4259415 & 1237664667891663313 & $20.692\pm0.004$ & $ 0.53077\pm0.00013$ & 0 &  ..
. &   $11.13^{+0.22}_{-0.18}$ &   ... \\
138.7072691+30.5814852 & 1237664668965601912 & $20.757\pm0.005$ & $ 0.66876\pm0.00013$ & 0 & 1.41 &   $10.86^{+0.26}_{-0.21}$ &   ... \\
138.7073207+30.5341867 & 1237664668965601842 & $20.743\pm0.005$ & $ 0.34103\pm0.00012$ & 0 & 1.23 &   $ 9.82^{+0.21}_{-0.19}$ &   ... \\
138.7099088+29.9916361 & 1237664668428600135 & $20.789\pm0.005$ & $ 0.53978\pm0.00014$ & 0 & 1.55 &   $10.83^{+0.16}_{-0.20}$ &   ... \\
138.7104152+29.4288178 & 1237664667891663325 & $20.707\pm0.005$ & $ 0.34555\pm0.00011$ & 0 & 1.31 &   $ 9.93^{+0.21}_{-0.20}$ &   ... \\
138.7114904+29.5393724 & 1237664667891663438 & $20.834\pm0.005$ & $ 0.60394\pm0.00011$ & 0 & 1.44 &   $11.12^{+0.12}_{-0.16}$ &   ... \\
138.7126167+30.2366575 & 1237664093439984562 & $20.645\pm0.005$ & $ 0.59847\pm0.00010$ & 0 & 1.12 &   $ 9.90^{+0.24}_{-0.20}$ &   ... \\
\enddata
\tablenotetext{a}{This table is available in its entirety in a machine-readable form in the online journal. A portion is shown here for guidance regarding its form and content. The full table contains 2994 galaxies all with an MMT Hectospec redshift. Among these galaxies, 2942 are outside the masked region.}
\tablenotetext{b}{SDSS ObjID from DR10. If DR10 ObjID is not available, we present SDSS DR7 ObjID that starts with `58'.}
\tablenotetext{c}{All redshifts are from this study.}
\tablenotetext{d}{(0) Outside masked regions; (1) Inside masked regions.}
\end{deluxetable}
\clearpage

\begin{deluxetable}{ccc}
\tabletypesize{\footnotesize}
\tablewidth{0pc} 
\tablecaption{Objects without redshifts at $R\leq20.6$ outside Masked Regions\tablenotemark{a}
\label{tab-samp_noz}}
\tablehead{
SHELS ID & SDSS ObjID\tablenotemark{b}  & $R$   \\
         &                              & (mag) 
}
\startdata
138.7025353+30.9340239 & 1237661381695505433 & $20.474\pm0.005$ \\
138.7040214+30.8569490 & 1237664669502538306 & $20.387\pm0.004$ \\
138.7049653+30.8566743 & 1237664669502538305 & $20.374\pm0.004$ \\
138.7089504+30.9843187 & 1237661381695570463 & $20.556\pm0.005$ \\
138.7111145+29.1171632 & 1237664880485466669 & $19.679\pm0.002$ \\
138.7189655+29.5848519 & 1237664667891662952 & $17.697\pm0.001$ \\
138.7233323+30.0294043 & 1237664668428665215 & $20.570\pm0.005$ \\
138.7255980+30.0542613 & 1237664668428665266 & $20.539\pm0.005$ \\
138.7257555+31.0038671 & 1237661381695570546 & $20.172\pm0.004$ \\
138.7264737+30.3206758 & 1237664093440049704 & $19.139\pm0.093$ \\
\enddata
\tablenotetext{a}{This table is available in its entirety in a machine-readable form in the onli
ne journal. A portion is shown here for guidance regarding its form and content.The full table contains 703 objects.}
\tablenotetext{b}{SDSS ObjID from DR10. If DR10 ObjID is not available, we include SDSS DR7 ObjID that starts with `58'.}
\end{deluxetable}

\clearpage

\begin{deluxetable}{cccccc}
\rotate
\tabletypesize{\footnotesize}
\tablewidth{0pc} 
\tablecaption{Data for the spectra in Figure 14\tablenotemark{a}
\label{tab-fig14}}
\tablehead{
 Wavelength ($\AA$) & S(11.00$-$11.50)\tablenotemark{b} & S(11.00$-$11.50) & S(11.00$-$11.50) & 
S(11.00$-$11.50) & S(10.50$-$11.00) \\
 & ($z=0.2-0.3$)\tablenotemark{c} & ($z=0.3-0.4$) & ($z=0.4-0.5$) & ($z=0.5-0.6$) & ($z=0.1-0.2$
) 
}
\startdata
3600.0 &  42.70 &  41.70 &  43.92 &  41.33 &  46.16 \\
3601.2 &  42.27 &  39.85 &  42.70 &  41.49 &  44.99 \\
3602.4 &  40.51 &  39.31 &  41.96 &  41.22 &  44.22 \\
3603.6 &  38.82 &  37.30 &  39.97 &  40.06 &  41.88 \\
3604.8 &  37.15 &  35.80 &  37.88 &  36.94 &  40.48 \\
3606.0 &  35.32 &  35.38 &  36.45 &  37.75 &  39.03 \\
3607.2 &  34.77 &  34.83 &  36.59 &  38.33 &  38.01 \\
3608.4 &  34.39 &  34.37 &  36.16 &  37.37 &  37.09 \\
3609.6 &  33.68 &  34.58 &  36.21 &  39.39 &  37.04 \\
3610.8 &  34.35 &  34.72 &  36.72 &  40.81 &  37.47 \\
\enddata
\tablenotetext{a}{This table is available in its entirety in  machine-readable form in the online journal. A portion is shown here for guidance
regarding its form and content. We show only 5 of the 22 columns of data. Each column corresponds to a spectrum in a panel of Figure 14.}
\tablenotetext{b}{Normalized Flux. Numbers in parentheses indicate the mass range for summed spectra.}
\tablenotetext{c}{Numbers in parentheses indicate the redshift range for summed spectra.}
\end{deluxetable}

\begin{deluxetable}{cccccc}
\rotate
\tabletypesize{\footnotesize}
\tablewidth{0pc} 
\tablecaption{Data for the spectra in Figure 15\tablenotemark{a}
\label{tab-fig15}}
\tablehead{
 Wavelength ($\AA$) & S( 9.08$-$ 9.32)\tablenotemark{b} & S( 9.32$-$ 9.47) & S( 9.47$-$ 9.62) & 
S( 9.62$-$ 9.75) & S( 9.75$-$ 9.86)
}
\startdata
3600.2 & 0.758 & 0.781 & 0.718 & 0.747 & 0.759 \\
3600.9 & 0.749 & 0.776 & 0.725 & 0.744 & 0.749 \\
3601.5 & 0.743 & 0.768 & 0.715 & 0.727 & 0.738 \\
3602.2 & 0.721 & 0.763 & 0.705 & 0.710 & 0.737 \\
3602.9 & 0.713 & 0.765 & 0.695 & 0.712 & 0.733 \\
3603.5 & 0.714 & 0.763 & 0.686 & 0.717 & 0.723 \\
3604.2 & 0.725 & 0.742 & 0.686 & 0.704 & 0.726 \\
3604.8 & 0.698 & 0.737 & 0.685 & 0.703 & 0.722 \\
3605.5 & 0.690 & 0.732 & 0.680 & 0.711 & 0.721 \\
3606.2 & 0.680 & 0.720 & 0.675 & 0.715 & 0.732 \\
\enddata
\tablenotetext{a}{This table is available in its entirety in machine-readable form in the online journal. A portion is shown here for guidance regarding its form and content. We show only 5 of the 10 columns of data. Each column corresponds to a spectrum in Figure 15.}
\tablenotetext{b}{Normalized Flux. Numbers in parentheses indicate the mass range for summed spectra.}
\end{deluxetable}

\clearpage

\clearpage

\clearpage

\clearpage

\clearpage


\clearpage

\clearpage

\clearpage

\clearpage

\clearpage

\clearpage


\clearpage

\clearpage

\clearpage

\clearpage

\clearpage

\clearpage

\end{document}